\documentclass[12pt]{reportj}
\usepackage{deluxetablej}
\usepackage{hyperref} %% Links all your references to pages
\makeatletter
\def\@to{to}
\makeatother
\usepackage{times}
\usepackage{graphicx}
\usepackage{tocloft}
\usepackage{fancyheadings}
\emergencystretch=\maxdimen
\usepackage{datetime}
\newdateformat{ddmonthyyyy}{\THEDAY\ \monthname[\THEMONTH]\ \THEYEAR}

\renewcommand\thesection{\arabic{section}}
\renewcommand\thesubsection{\thesection.\arabic{subsection}}

\def\ssection#1{\setcounter{subsection}{0} \refstepcounter{section} \section*{\hbox to \hsize{\large\bf \arabic{section}. #1\hfill }}\label{sec} \addcontentsline{toc}{section}{\arabic{section}. #1}}
\def\ssubsection#1{\setcounter{subsubsection}{0} \refstepcounter{subsection}\subsection*{\hbox to \hsize{\normalsize\bfseries\itshape \arabic{section}.\arabic{subsection} #1\hfill}}\label{subsec} \addcontentsline{toc}{subsection}{\arabic{section}.\arabic{subsection} #1}}
\def\ssubsubsection#1{\refstepcounter{subsubsection}\subsection*{\hbox to \hsize{\normalsize\it \arabic{section}.\arabic{subsection}.\arabic{subsubsection} #1\hfill}}\label{subsubsec} \addcontentsline{toc}{subsubsection}{\arabic{section}.\arabic{subsection}.\arabic{subsubsection} #1}}

\def\ssectionstar#1{\section*{\hbox to \hsize{\large\bf #1\hfill}} \addcontentsline{toc}{section}{#1}}
\def\ssubsectionstar#1{\subsection*{\hbox to \hsize{\normalsize\bfseries\itshape #1\hfill}} \addcontentsline{toc}{subsection}{#1}}
\def\ssubsubsectionstar#1{\subsection*{\hbox to \hsize{\normalsize\it  #1\hfill}} \addcontentsline{toc}{subsection}{#1}}

\renewcommand{\cftaftertoctitle}{%
\mbox{}\hfill{\normalfont Page}}
\begin{document}

~\\

% ST Logo in the top left
\vspace{-2.4cm}
\noindent\includegraphics*[width=0.295\linewidth]{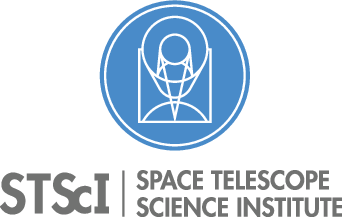}

\vspace{-0.4cm}

\begin{flushright}
    %%% Put the instrument, year, and ISR number here (and also below) %%%
    {\bf Instrument Science Report STIS 2025-01(v1)}
    
    \vspace{1.1cm}
    
    %%% Put ISR Title %%%
    {\bf\Huge STIS Cycle 30 Calibration Programs}
    
    \rule{0.25\linewidth}{0.5pt}
    
    \vspace{0.5cm}
    
    %Put Authors
    D. Welty$^1$, R. Bohlin$^1$, J. Carlberg$^1$, M. Dallas$^1$, S. Hernandez$^{1,2}$, \\
    A. Jones$^1$, S. Lockwood$^1$, S. Medallon$^1$, E. Rickman$^{1,3}$, D. Stapleton$^1$, \\
    and T. Wheeler$^1$ 
    \linebreak
    \newline
    %Put Author's affiliations
    \footnotesize{$^1$ Space Telescope Science Institute, Baltimore, MD\\
                  $^2$ AURA for the European Space Agency\\
                  $^3$ European Space Agency (ESA), ESA Office, Space Telescope Science Institute\\}
    
    \vspace{0.5cm}
    
    % Date in DD Month YYYY format based on when you compile it
     21 January 2025
\end{flushright}

\vspace{0.1cm}

%%% ABSTRACT %%%
\noindent\rule{\linewidth}{1.0pt}
\noindent{\bf A{\footnotesize BSTRACT}}

{\it \noindent 
We discuss the suite of STIS calibration programs executed during HST Cycle 30, covering the period 2022~Nov~07 through 2023~Nov~05.
For each of the 19 current regular calibration programs, we provide brief descriptions of the objectives, observations, analysis procedures, and results -- with comparisons to the results from previous cycles and to desired accuracies, as well as references to more detailed analyses of the calibration data.
Many of these calibration programs produce routine reference file deliveries or demonstrate the continuing applicability of existing reference files for processing STIS observations.
This ISR provides a brief snapshot of the current instrument performance, similar to those given in annual reports for Cycles 7--10 and 17--21.
Two Appendices briefly discuss the state of the onboard calibration lamps and the ongoing major effort to revise the flux calibration for the many STIS spectroscopic and imaging modes.}

\vspace{-0.1cm}
\noindent\rule{\linewidth}{1.0pt}

%% Table of Contents
%% Need to compile twice to get the page numbers correct
\renewcommand{\cftaftertoctitle}{\thispagestyle{fancy}}
\tableofcontents

%%% MAIN TEXT BELOW %%%

%\setlength{\textwidth}{170mm}

\vspace{-0.3cm}
\ssectionstar{Introduction}\label{sec:Introduction}

During each Hubble Space Telescope (HST) cycle, the Space Telescope Imaging Spectrograph (STIS) team executes a set of regular calibration programs designed to monitor the health and performance of the three STIS detectors and other instrumental components.
Special calibration programs, designed to address specific issues, may also be implemented during individual cycles.
The quantitative data from these programs are used to generate reference files characterizing the instrumental behavior, which are used for both {\sc calstis} pipeline reductions of STIS science data and for predicting instrumental performance with the STIS Exposure Time Calculator (ETC). 
While the behavior of STIS is at this point fairly stable and well understood, some characteristics have been evolving with time -- e.g., detector sensitivity (as a function of wavelength), dark count rates, wavelength calibration, detector cosmetics, the brightness of the calibration lamps, and other, more subtle properties.  
The behavior of STIS can also be affected if the instrument is shut down for more than several days (e.g., during an extended ``safing'' of the observatory).
Such changes can be seen both in data obtained for the calibration programs and in science data obtained and analyzed by guest observers.
This document provides a brief description of the current state of the instrument, based primarily on analyses of the calibration observations carried out during HST Cycle 30, nominally covering the period 2022~Nov through 2023~Oct (though more recent results are noted in a few cases). 

The current suite of regular (generally annual) calibration programs is also fairly stable.
While the current set of programs is not as extensive as those executed during the original commissioning period (Cycle 7) or during the re-commissioning following the repair of STIS in Servicing Mission 4 (Cycle 17; \href{https://www.stsci.edu/files/live/sites/www/files/home/hst/instrumentation/stis/documentation/instrument-science-reports/_documents/2013_04.pdf}{STIS ISR 2013-04}), a number of the programs that have continued are fairly similar to their predecessors.
In a number of cases, minor changes have been made over the years -- to adjust for decreasing detector sensitivities or the fading of the internal calibration lamps, to monitor additional instrumental characteristics, or (in some cases) to reduce the number of specific modes that are monitored.
In Cycle 30, a program to monitor the three primary white dwarf flux standard stars -- which had executed as a special calibration program in several previous cycles -- was added to the regular suite, to be run every other cycle.
No special calibration programs were executed during Cycle 30.
There were only three short ($\sim$1 day) ``safing'' events during Cycle 30 (2023~Aug~08, 2023~Aug~12, 2023~Aug~28), none of which lasted long enough to significantly affect the behavior of the STIS detectors.

The 19 programs in the current calibration suite are listed in Table~\ref{tab:progs}, with the Cycle 30 PI, the number of external and internal orbits used, and an indication of the frequency of execution.
Only five of the programs obtain spectroscopic or imaging data for external targets -- either individual flux standard stars or standard stellar fields.
The other 14 programs execute in ``internal'' orbits -- either observing one of the onboard calibration lamps or no source at all (e.g., observations of the bias and dark rates).
The entire Cycle 30 STIS calibration program used 1342 internal orbits and 26 external orbits.

Brief discussions of several issues related to the STIS calibration program are presented in two Appendices.
Appendix A provides some information regarding the current state of the onboard calibration lamps.
Appendix B describes the extensive effort underway to refine the flux calibration of the various STIS spectroscopic and imaging modes, in response to recent improvements in stellar models.
For more details regarding particular calibration programs and issues, users may consult the following resources: 

\href{https://hst-docs.stsci.edu/stisihb}{STIS Instrument Handbook} 

\href{https://hst-docs.stsci.edu/stisdhb}{STIS Data Handbook} 

\href{https://www.stsci.edu/hst/instrumentation/stis/documentation/instrument-science-reports}{STIS Instrument Science Reports} -- some listed in the references (below)

\href{https://www.stsci.edu/hst/instrumentation/stis/calibration}{STIS calibrations page} -- and the individual phase II proposals linked there 

\href{https://www.stsci.edu/hst/instrumentation/stis/performance}{STIS performance web page} -- instrumental characteristics and monitoring

\href{https://hst-crds.stsci.edu/}{HST Calibration Reference Data System (CRDS)} -- reference files

\href{https://stsci.service-now.com/hst}{HST HelpDesk}

%!!!!!!!!!!!!!!!!!!!!!!!!!!!!!!!!!!!!!!!!!!!!!!!!!!!!!!!!!!!!!!!!!!!!!!!!!!!!!!!!!!!!!!!!!!!!!!!!!!!!!!!!!!!!!!!!!!!!!!!!!!!!!!!!!!!!!!!!!!!!!!!!!!!!!!!!!!!!!!!!!!!!!!!!!!!!!!!!!!!!!!!!!!!!!!!!!!!!!!!!!!!!!!!!!!!!!!!!!!!!!!!!!!!!!!!!!!!!!!!!
%THIS SECTION NEEDS TO GO AFTER THE END OF THE FIRST PAGE AND BEFORE THE END OF THE SECOND PAGE
%Fill in Instrument, Year, and ISR Number and delete "newpage" immediately after this message
%!!!!!!!!!!!!!!!!!!!!!!!!!!!!!!!!!!!!!!!!!!!!!!!!!!!!!!!!!!!!!!!!!!!!!!!!!!!!!!!!!!!!!!!!!!!!!!!!!!!!!!!!!!!!!!!!!!!!!!!!!!!!!!!!!!!!!!!!!!!!!!!!!!!!!!!!!!!!!!!!!!!!!!!!!!!!!!!!!!!!!!!!!!!!!!!!!!!!!!!!!!!!!!!!!!!!!!!!!!!!!!!!!!!!!!!!!!!!!!
%\newpage

\lhead{}
\rhead{}
\cfoot{\rm {\hspace{-1.9cm} Instrument Science Report STIS 2025-01(v1) Page \thepage}}
%%%%%%%%%%%%%%%%

\clearpage

\begin{deluxetable}{cllccc}
    \tabcolsep 4pt
    \tablewidth{0pt}
    \tablecaption{STIS Cycle 30 Calibration Programs \label{tab:progs}}
    \tabletypesize{\footnotesize}
    % Heading labels
    \tablehead{
      \colhead{PID} &  \colhead{Program} & \colhead{PI}  & \colhead{Ext Orbits} & \colhead{Int Orbits} & \colhead{Frequency} }
    \startdata

16945 & CCD Performance Monitor                          & S. Lockwood  &  0 & 14 & 2x/yr \\
16946-48 & CCD Dark Monitor                              & S. Medallon  &  0 &728 & 2x/day \\
16949-50 & CCD Bias and Read Noise Monitor               & S. Lockwood  &  0 &364 & daily \\
16951 & CCD Hot Pixel Annealing                          & S. Medallon  &  0 & 39 & 13x/yr \\
16944 & CCD Spectroscopic Flat Field Monitor             & J. Carlberg  &  0 & 19 & ~monthly \\
16952 & CCD Imaging Flat Field Monitor                   & D. Welty     &  0 &  4 & 4x/yr \\
16954 & CCD Spectroscopic Dispersion Solution Monitor    & D. Welty     &  0 &  3 & 1x/yr \\
16953 & CCD Sparse Field CTE                             & S. Lockwood  &  0 & 50 & 1x/yr \\
16955 & CCD Full Field Sensitivity Monitor               & S. Lockwood  &  1 &  0 & 1x/yr \\
16957 & CCD Spectroscopic Sensitivity Monitor            & S. Hernandez &  5 &  0 & 4x/yr \\
16958 & Slit Wheel Repeatability                         & A. Jones     &  0 &  1 & 1x/yr \\
16959 & MAMA Spectroscopic Dispersion Monitor            & D. Welty     &  0 &  7 & 1x/yr \\
16956 & MAMA Full Field Sensitivity Monitor              & S. Lockwood  &  3 &  0 & 1x/yr \\
16960 & MAMA Spectroscopic Sensitivity and Focus Monitor & S. Hernandez & 12 &  0 & 8x/yr \\
16961 & FUV MAMA Dark Monitor                            & S. Lockwood  &  0 & 48 & 9x/yr \\
16962 & NUV MAMA Dark Monitor                            & A. Jones     &  0 & 52 & bi-weekly \\
16963 & FUV MAMA Flat Field Monitor                      & A. Jones     &  0 & 11 & 11x/yr \\
16964 & MAMA Fold Distribution                           & T. Wheeler   &  0 &  2 & 1x/yr \\
16966 & Monitoring the Three Primary WD Standard Stars   & R. Bohlin    &  5 &  0 & 1x/yr \\
\hline
total &                                                  &              & 26 &1342& \\
%                 \hline \vspace{-0.3cm} \\    
    \enddata
\end{deluxetable}

%%% -~-~-~-~-~-~-~-~-~- CCD Performance -~-~-~-~-~-~-~-~-~- %%%

\clearpage
%\vspace{-0.3cm}
\ssectionstar{CCD Performance Monitor (16945; S. Lockwood / M. Dallas)}\label{sec:sec_cperf}

{\bf Objectives:}
This program monitors the overall baseline performance of the STIS CCD.

\noindent
{\bf Observations:}
Full frame bias and flat field (imaging and spectroscopic) exposures are taken in order to measure read noise, charge transfer efficiency (CTE), spurious charge, and gain.
Different apertures and exposure times are used for the flat fields; gain=1,2,4,8 are used for the imaging flats and the bias exposures.  
Bias frames are also taken in subarrays to check the bias level for ACQ and ACQ/PEAK observations.  
All exposures use amplifier D.
The data were obtained in two sets of seven internal orbits, in March and September of 2023.
 
\noindent
{\bf Analysis:} 
Determinations of the read noise and CTE are discussed below, under the more specific programs monitoring those effects.
More detailed discussions of the analysis of CCD performance data may be found in \href{https://www.stsci.edu/files/live/sites/www/files/home/hst/instrumentation/stis/documentation/instrument-science-reports/_documents/200902.pdf}{STIS ISR 2009-02}, which describes the behavior of the STIS CCD just after SM-4, and \href{https://www.stsci.edu/files/live/sites/www/files/home/hst/instrumentation/stis/documentation/instrument-science-reports/_documents/2017_05.pdf}{STIS ISR 2017-05}.

\begin{figure}[!hb]
  \centering
  \includegraphics[width=120mm]{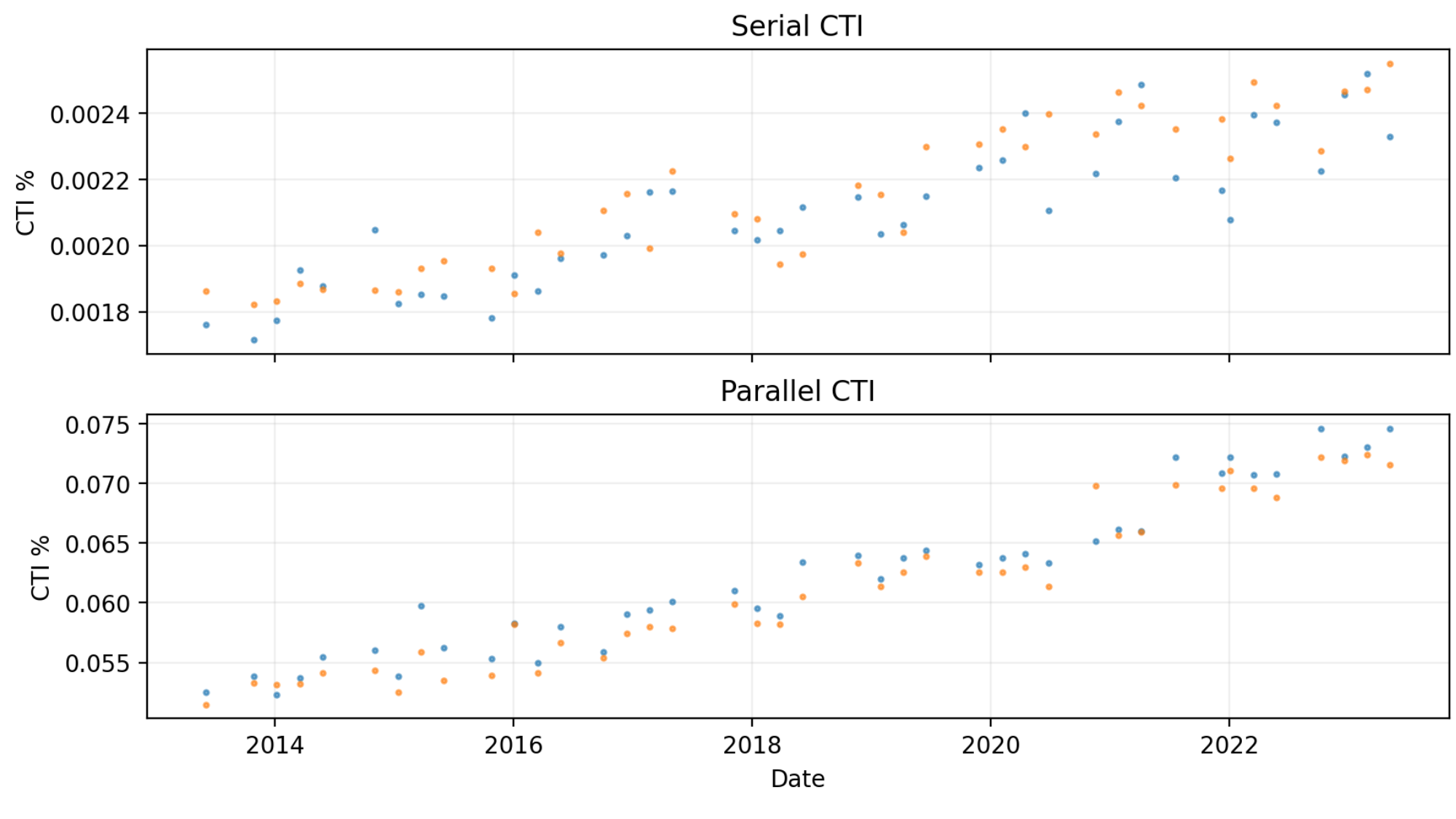}
    \caption{Charge transfer inefficiency (CTI = 1 - CTE) for the STIS CCD.  The two panels show the general increases in the serial and parallel CTI, for the 0.3s (orange) and 3.6s (blue) exposures at gain=4, since SM4.}
    \label{fig:ccti}
\end{figure}

\begin{figure}[!ht]
  \centering
  \includegraphics[width=160mm]{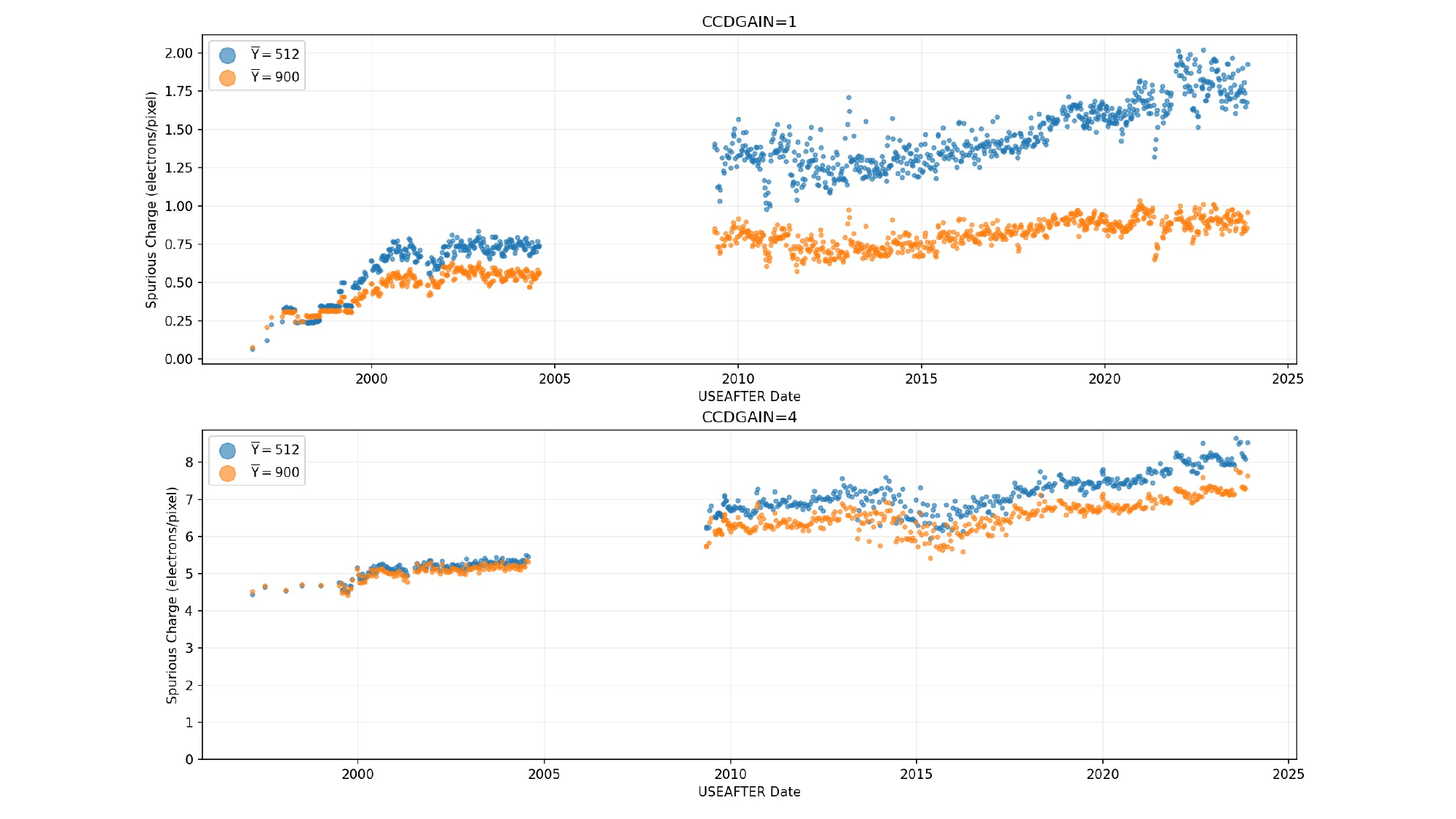}
    \caption{Spurious charge in the reference superbiases, for gain=1 (top) and gain=4 (bottom).  The values at pseudo-aperture E1 (orange) are somewhat lower than those at the center of the CCD (blue).}
    \label{fig:spchrg}
\end{figure}

\begin{figure}[!hb]
  \centering
  \includegraphics[width=150mm]{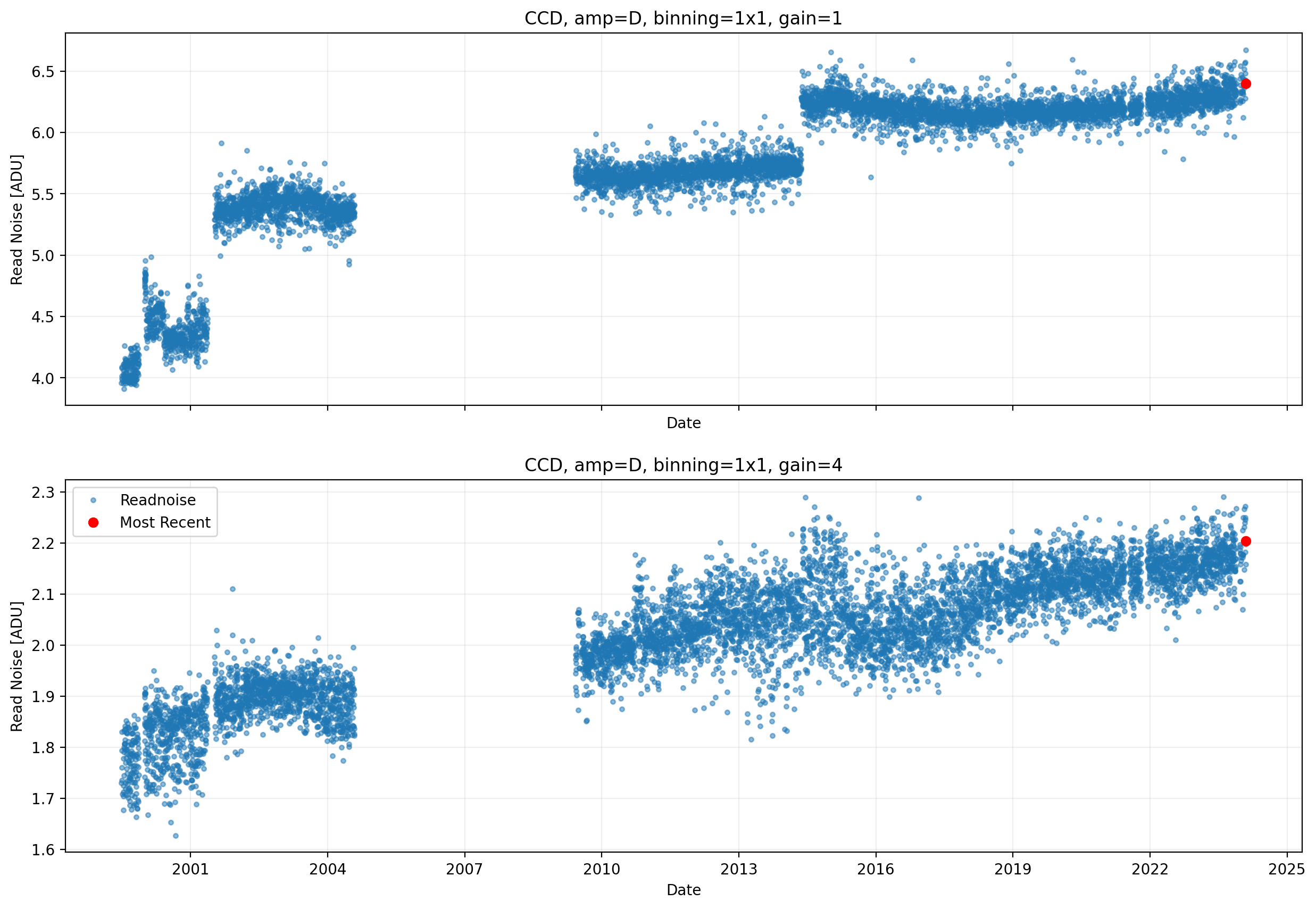}
    \caption{CCD read noise for gain=1 (top) and gain=4 (bottom).  The read noise is consistent with the recent slow increases at both gains.  The right-most red point is the most recent measurement.}
    \label{fig:crn}
\end{figure}

\noindent
{\bf Results:}
Up-to-date plots of the CCD housing temperature (from engineering telemetry) and read noise are available on the main \href{https://www.stsci.edu/hst/instrumentation/stis/performance/monitoring}{STIS monitors} web pages.
Plots of the CTE, spurious charge, and various other quantities are available at \href{https://www.stsci.edu/\~STIS/monitors}{https://www.stsci.edu/$\sim$STIS/monitors}.
Overall, the various measures of CCD behavior remain consistent with recent trends.
The CCD housing temperature (thought to be the best currently available proxy for the detector temperature) is typically of order 21$\pm$2 K.
Both parallel and serial charge transfer inefficiency (CTI = 1 - CTE) continue their slow increase (Fig.~\ref{fig:ccti}); the current CTI values, as measured by the Extended Pixel Edge Response (EPER) method, are $\sim$0.073\% and $\sim$0.0024\%, respectively.
``Spurious charge'' refers to structure along the CCD columns, thought to be due to charge leakage from hot pixels -- thus increasing as the CCD accumulates radiation damage.
At the central row in the bias frames, the spurious charge is currently $\sim$1.8 e$^-$/pix for gain=1 and $\sim$8.3 e$^-$/pix for gain=4, and is somewhat lower at the E1 pseudo-apertures near row 900 (Fig.~\ref{fig:spchrg}).
The read noise, for amp D, is $\sim$6.3 ADU for gain=1 and $\sim$2.17 ADU ($\sim$8.7 e$^-$) for gain=4 (Fig.~\ref{fig:crn}).
The bias level, read noise, and gain may be used to update the CCDTAB reference file (last done in 2017 Jun).

%%% -~-~-~-~-~-~-~-~-~- CCD Darks -~-~-~-~-~-~-~-~-~- %%%

\clearpage
%\vspace{-0.3cm}
\ssectionstar{CCD Dark Monitor (16946-16948); S. Medallon / D. Welty)}\label{sec:sec_cdark}

{\bf Objectives:}
This program monitors the dark current for the STIS CCD and is also used to track hot and bad CCD pixels.

\noindent
{\bf Observations:}
Two internal 1-orbit visits are executed each day; each visit obtains two 60-sec and one 1100-sec dark exposures (for amp D and gain=1), bracketing the exposure times for many of the science exposures taken with the CCD.

\noindent
{\bf Analysis:}
The individual exposures are examined for consistency, cleaned of cosmic rays, adjusted for the temperature of the side-2 electronics, then combined into weekly ``superdarks'', which are delivered to CRDS as reference files.
The RMS deviations in the superdarks should be $<$ 0.012 cts/s/pix.
The weekly reference files are used in the data reduction pipeline to correct contemporaneous raw CCD science exposures for the dark current.

\begin{figure}[!hb]
  \centering
  \includegraphics[width=150mm]{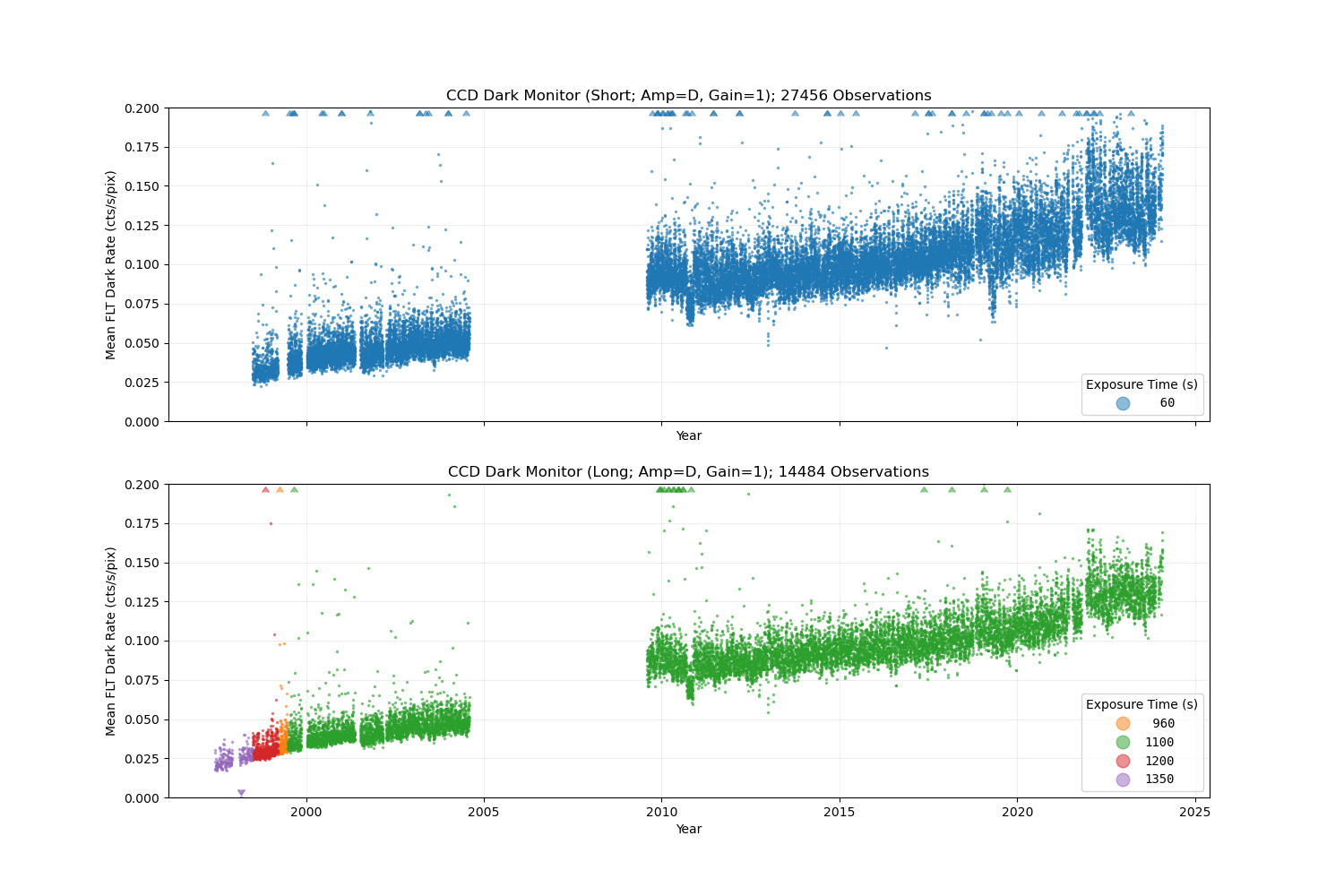}
    \caption{CCD dark count rates for short (60-sec; top) and long (1100-sec; bottom) exposures.
             During Cycle 30, the dark rates remained relatively stable, with an average around 0.13 cts/s/pix.}
    \label{fig:cdark}
\end{figure}

\noindent
{\bf Results:}
In addition to the superdark reference file delivery, the dark counts are posted weekly to the \href{https://www.stsci.edu/hst/instrumentation/stis/performance/monitoring}{STIS monitors} web page, and the yearly averages are incorporated into the STIS Exposure Time Calculator (for S/N calculations). 
During Cycle 30, the dark rates remained relatively stable, with an average around 0.13 cts/s/pix (Fig.~\ref{fig:cdark}).

%%% -~-~-~-~-~-~-~-~-~- Bias and Read Noise -~-~-~-~-~-~-~-~-~- %%%

\clearpage
%\vspace{-0.3cm}
\ssectionstar{CCD Bias and Read Noise Monitor (16949-16950; S. Lockwood / D. Stapleton)}\label{sec:sec_cbrn}

{\bf Objectives:}
This program monitors the bias at gain=1 and gain=4 to construct superbias frames, track hot columns, and measure the read noise.

\noindent
{\bf Observations:}
Full frame bias exposures (14 at gain=1, 3 at gain=4) with amp D, plus several exposures with amps A and C, are obtained in one internal orbit each day.
 
\noindent
{\bf Analysis:} 
For each gain, the read noise is calculated from the differences between individual bias images (to remove the effects of structure in the bias images). 
Superbias image statistics are compared with previous values.

\noindent
{\bf Results:}
Weekly superbias reference files are generated; the superbiases should have RMS $<$ 0.95 e$^-$ for gain=1 and RMS $<$ 1.13 e$^-$ for gain=4 (1x1 binning).
The read noise values are posted to the \href{https://www.stsci.edu/hst/instrumentation/stis/performance/monitoring}{STIS monitors} web page.
The read noise remains consistent with recent trends (Fig.~\ref{fig:crn}); it is currently $\sim$6.3 ADU for gain=1 and $\sim$2.17 ADU ($\sim$8.7 e$^-$) for gain=4.

%%% -~-~-~-~-~-~-~-~-~- Annealing -~-~-~-~-~-~-~-~-~- %%%

\clearpage
%\vspace{-0.3cm}
\ssectionstar{CCD Hot Pixel Annealing (16951; S. Medallon / D. Welty)}\label{sec:sec_canneal}

{\bf Objectives:}
This program anneals the CCD to try to limit the increase in hot pixels.
Pre- and post-anneal bias, dark, and flat field exposures are obtained to assess the effectiveness of the annealing and to track various aspects of the behavior of the CCD.

\noindent
{\bf Observations:}
Every four weeks, three internal orbits are used to (1) obtain a series of pre-anneal bias, dark, and flat field exposures; (2) allow the CCD to warm from $\sim$$-$83 C to $\sim$+5 C (over $\sim$12 hr) by turning off the thermoelectric cooler; (3) return the CCD to its normal operating temperature (over $\sim$4 hr) by turning the cooler back on; and (4) obtain a corresponding post-anneal set of bias, dark, and flat field exposures.
All 13 executions of the anneal during Cycle 30 were successful.

\noindent
{\bf Analysis:}
Comparison of the pre- and post-anneal data -- e.g., the locations, brightnesses, and number of hot pixels -- indicates which hot pixels have been ``fixed'' or ``reset'' by the annealing, which are persistent, and which of the latter were affected by the annealing.
From those comparisons, estimates are made for the growth rate of hot pixels, the effectiveness of the annealing, and the mean and median dark count rates.
The latter can be compared with contemporaneous values from the daily CCD dark monitor.
Examination of the flat fields can reveal any contamination of the detector window that might have occured during the annealing. 

\begin{figure}[!hb]
  \centering
  \includegraphics[width=150mm]{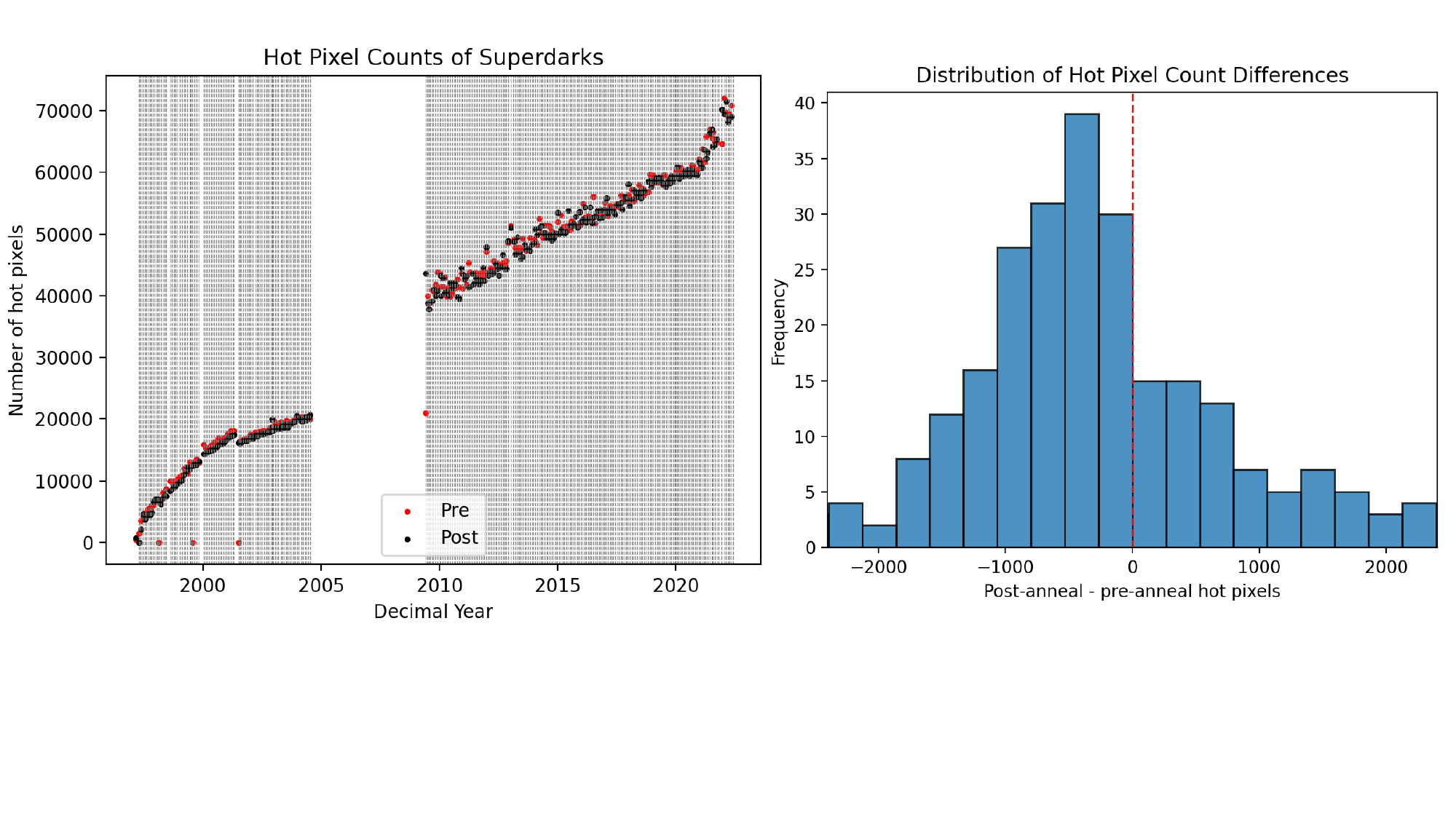}
    \caption{({\it left}) Comparison of the number of hot pixels ($>$ 0.1 e$^-$/s) in STIS CCD superdarks before (red) and after (black) annealing.  ({\it right}) Histogram of the difference in the number of hot pixels -- after anneal minus before anneal -- for post-SM4 anneals.  While there appear to be many persistent hot pixels, the anneals do yield some reduction in the number, on average.}
    \label{fig:anneal}
\end{figure}

\noindent
{\bf Results:}
While there appears to be an increasing number of persistent hot pixels (e.g., \href{https://www.stsci.edu/files/live/sites/www/files/home/hst/instrumentation/stis/documentation/instrument-science-reports/_documents/2017_05.pdf}{STIS ISR 2017-05}), the anneals do yield some reduction, on average (Fig.~\ref{fig:anneal}).  
That reduction is seen most easily in comparisons of superdarks, where the effects of cosmic rays have been more effectively removed.

%%% -~-~-~-~-~-~-~-~-~- CCD Spectroscopic Flats -~-~-~-~-~-~-~-~-~- %%%

\clearpage
%\vspace{-0.3cm}
\ssectionstar{CCD Spectroscopic Flat Field Monitor (16944; J. Carlberg / M. Dallas)}\label{sec:sec_cspfl}

{\bf Objectives:}
This program obtains flat field exposures with a medium-resolution grating to monitor the pixel-to-pixel sensitivity variations characterizing spectroscopic CCD observations.
If significant changes are noted, the data can be used to generate new CCD p-flats. 

\noindent
{\bf Observations:}
Two sets of exposures are obtained -- all using the tungsten lamp, gain=4, and the G430M/5126 setting.
Nine 1-orbit visits, taken every 30-45 days, each obtain sets of nine 21-sec exposures through the clear 50CCD aperture.
Ten 1-orbit visits, closer in time, each obtain pairs of 550-sec exposures through the 52x2 aperture, at five different slit offset positions (in order to uniformly illuminate the detector by shifting the locations of the occulting bars).

\noindent
{\bf Analysis:}
Initial analysis involves comparing the overall count rates with those seen in previous cycles, as a quick check for lamp degradation or observing problems, and visually inspecting the positions of the many dust motes.
Cycle 30 p-flats can be created using the python code from STIS TIR 2013-02\footnotemark, then compared with the current pipeline p-flats (from Cycle 19) and those from intermediate cycles to determine if new p-flats should be delivered. 
\footnotetext{This reference may be obtained by request through the \href{https://stsci.service-now.com/hst}{HST HelpDesk}.}

\begin{figure}[!hb]
  \centering
  \includegraphics[width=85mm]{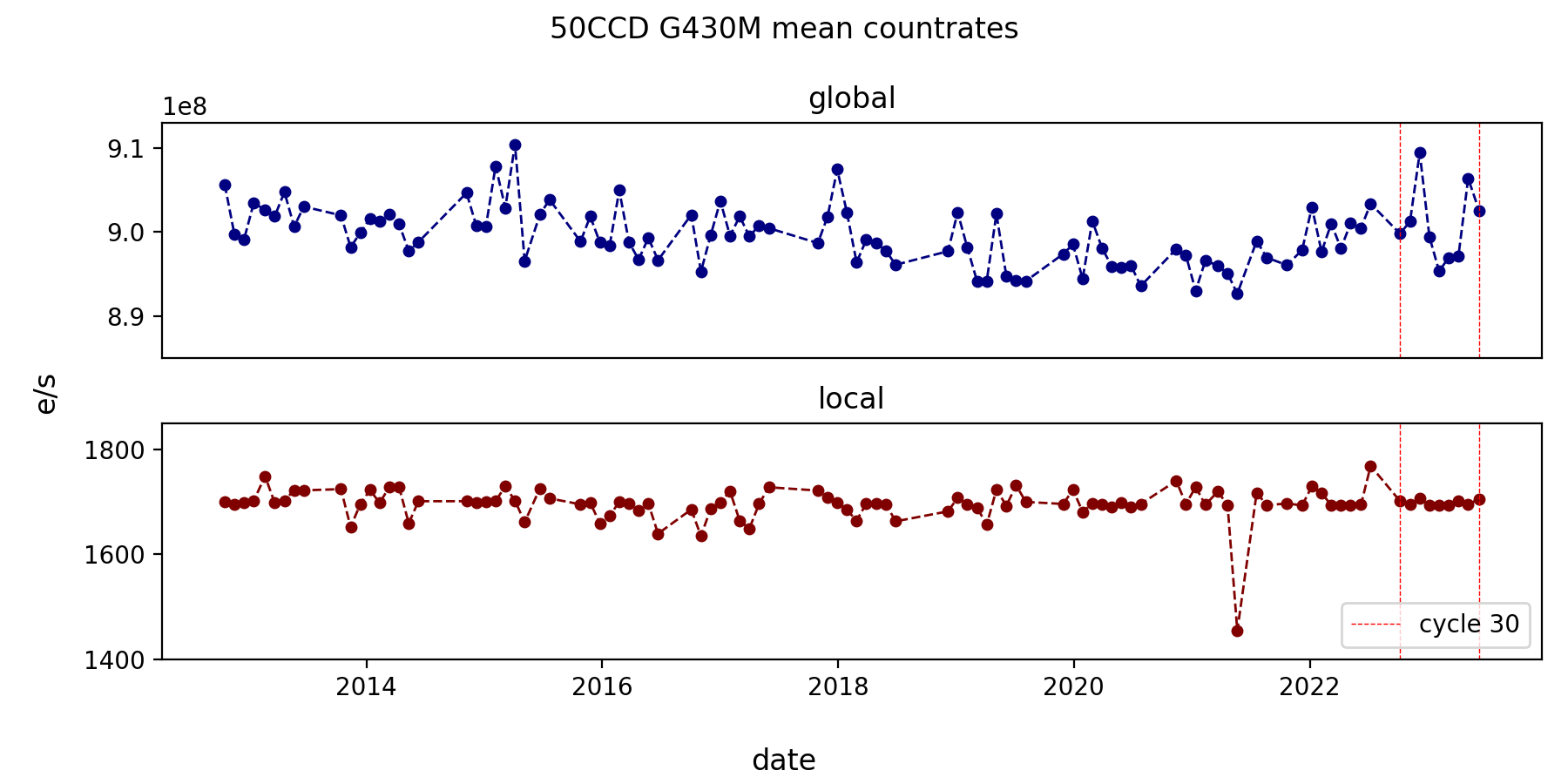}
  \includegraphics[width=85mm]{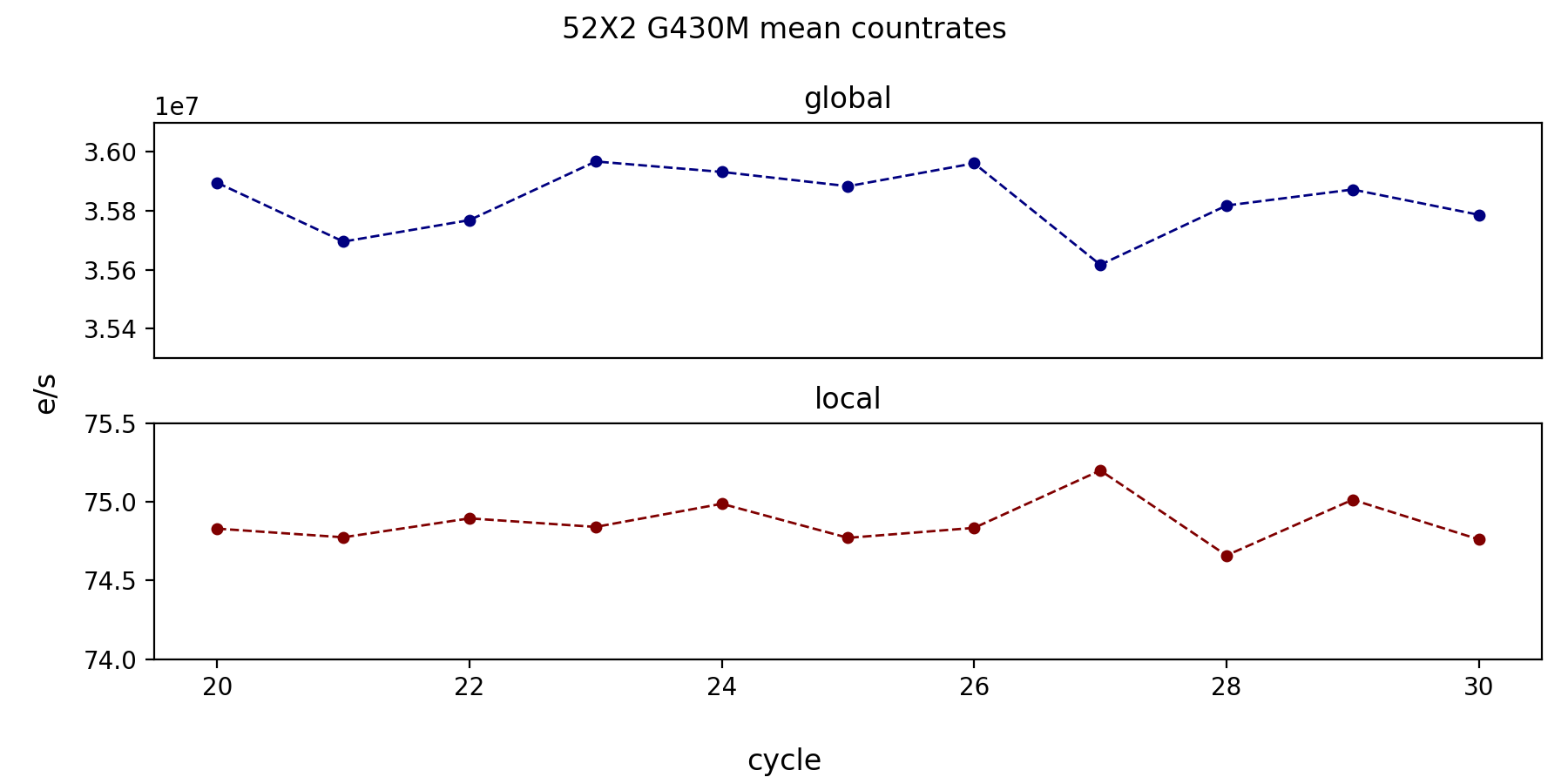}
    \caption{Mean global and local (brightest pixel) count rates for the 50CCD (top) and 52x2 (bottom) spectroscopic flat fields.  The Cycle 30 rates are consistent with those seen in prior cycles.}
    \label{fig:cspflat}
\end{figure}

\noindent
{\bf Results:}
The local (brightest pixel) and global count rates remain consistent with past programs for both the 50CCD and 52x2 exposures (Fig.~\ref{fig:cspflat}). 
There is no evidence for significant evolution of the dust motes and other small-scale features. 
New pipeline p-flats are not expected to be needed at this time, based on analyses of p-flats through Cycle 24 that found only marginal improvement when using the more contemporary flats.

%%% -~-~-~-~-~-~-~-~-~- CCD Imaging Flats -~-~-~-~-~-~-~-~-~- %%%

\clearpage
%\vspace{-0.3cm}
\ssectionstar{CCD Imaging Flat Field Monitor (16952; D. Welty)}\label{sec:sec_cimfl}

{\bf Objectives:}
This program obtains high-S/N white light imaging flat field exposures -- for monitoring and to enable new reference pixel-to-pixel flats (p-flats) for imaging and/or coronagraphic observations.

\noindent
{\bf Observations:}
Every 3 months, a series of exposures is obtained with the tungsten lamp -- 12 0.3-sec exposures through the clear 50CCD aperture and one 0.3-sec exposure through the 50CORON aperture.
Combining all 48 50CCD exposures should yield an average signal of about 620000 ADU/pix (as in past cycles) -- sufficient for the creation of a high accuracy ($\sim$1\%) imaging p-flat.
The 50CORON observations can be used to monitor the stability of observations using that aperture.

\noindent
{\bf Analysis:}
For each visit, the individual images are compared -- count rates in different parts of the image, number and location of dust motes.
After removing cosmic rays, the images can be combined to produce a visit-level super-flat; the four visit-level super-flats can then be combined to produce a cycle-level super-flat.
Removal of low-frequency variations, via spline fites to the rows and columns, then yields a cycle-level p-flat that may be compared with the p-flats produced in previous cycles -- and with the pre-launch p-flat that is currently used in the pipeline. 

\noindent
{\bf Results:}
Comparison of the count rates, in nine regions spread over the images, indicates that there are variations at the several percent level between the images in a given visit, due to the heating/cooling of the lamp during/between the exposures.
The count rates for the visit-level summed images, however, agree within 1\% -- both visit to visit and cycle to cycle (over the past several cycles) -- though there does appear to be a slight increase of about 0.3\% per year over the last 2-3 cycles.
The position, number, and appearance of the dust motes appears to be stable, apart from the slight rotation of the format discussed in \href{https://www.stsci.edu/files/live/sites/www/files/home/hst/instrumentation/stis/documentation/instrument-science-reports/_documents/2022_01.pdf}{STIS ISR 2022-01}.
We note, however, that a full analysis of the imaging flat data, and construction of new p-flats, has not been performed for many cycles.

\begin{figure}[!hb]
  \centering
  \includegraphics[width=75mm]{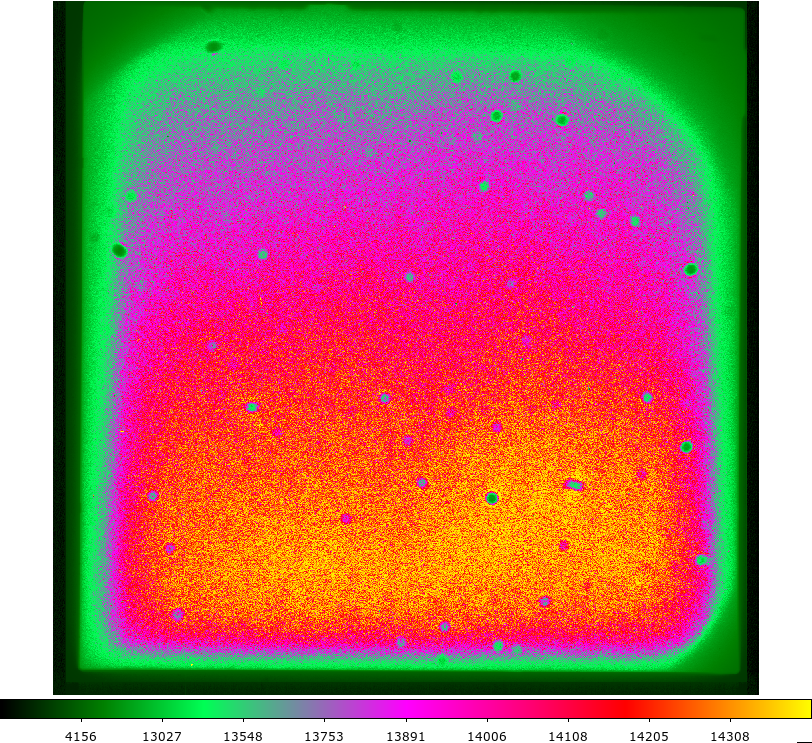}
    \caption{Sample raw CCD imaging flat field from Cycle 30, showing large-scale variations in count rates over the image and the many dust motes.}
    \label{fig:cimflat}
\end{figure}

%%% -~-~-~-~-~-~-~-~-~- Sparse Field CTE -~-~-~-~-~-~-~-~-~- %%%

\clearpage
%\vspace{-0.3cm}
\ssectionstar{CCD Sparse Field CTE (16953; S. Lockwood)}\label{sec:sec_ccte}

{\bf Objectives:}
This program measures Charge Transfer Inefficiency (CTI = 1 - CTE) in a spectroscopic-like configuration, using cross-dispersed tungsten lamp exposures at various Y-positions and flux levels.
Correction for the effects of CTI is required for obtaining accurate positional information and spectrophotometric fluxes from STIS CCD exposures -- particularly for faint targets.
 
\noindent
{\bf Observations:}
Cross-dispersed tungsten lamp flats are taken in the following configurations to yield 6 flux levels: G430M/0.05X31NDA ({0.3, 0.6, 0.9, 2.3} s), MIRVIS/0.05X31NDA (0.3 s), MIRVIS/0.05X31NDB (0.3 s).
The ``POS='' restricted optional parameter is used to represent specific Mode Select Mechanism values, stepping across 5 y-positions symmetric about the center of the detector. 
The observations are read out in alternately opposite parallel directions, using amplifier pairs with the same serial readout direction (A/C or B/D).
While routine STIS observations use amp D due to its lower read noise, the degradation of the symmetric amp B after the switch to side B electronics in 2009 (\href{https://www.stsci.edu/files/live/sites/www/files/home/hst/instrumentation/stis/documentation/instrument-science-reports/_documents/200902.pdf}{STIS ISR 2009-02}) now precludes performing this test on amp D.
The amp A/C pair is therefore used, as a proxy for the amp B/D pair.
Biases and darks are taken with the alternate amplifiers to reduce the lamp data.
Annual programs now alternate between gain=1 and gain=4 observations; gain=4 was used in Cycle 30.
All 50 visits executed successfully, during the first third of the cycle.

\noindent
{\bf Analysis:} 
CTI measurements are made using an ``internal sparse field test'', in which images are read out in either direction along both parallel and serial registers.
The further the charge needs to be shifted to be read out (in a given direction), the more charge is lost.
Comparison of the charge read out using amps A and C, using a legacy IDL code, thus yields a measure of the CTI as a function of source flux.
This analysis currently does not include the contribution of the spurious charge to the detector background level in the calculation of the CTI.
\href{https://www.stsci.edu/files/live/sites/www/files/home/hst/instrumentation/stis/documentation/instrument-science-reports/_documents/200902.pdf}{STIS ISR 2009-02} describes the derivation of the CTE coefficients currently used in the pipeline.

\begin{figure}[!hb]
  \centering
  \includegraphics[width=120mm]{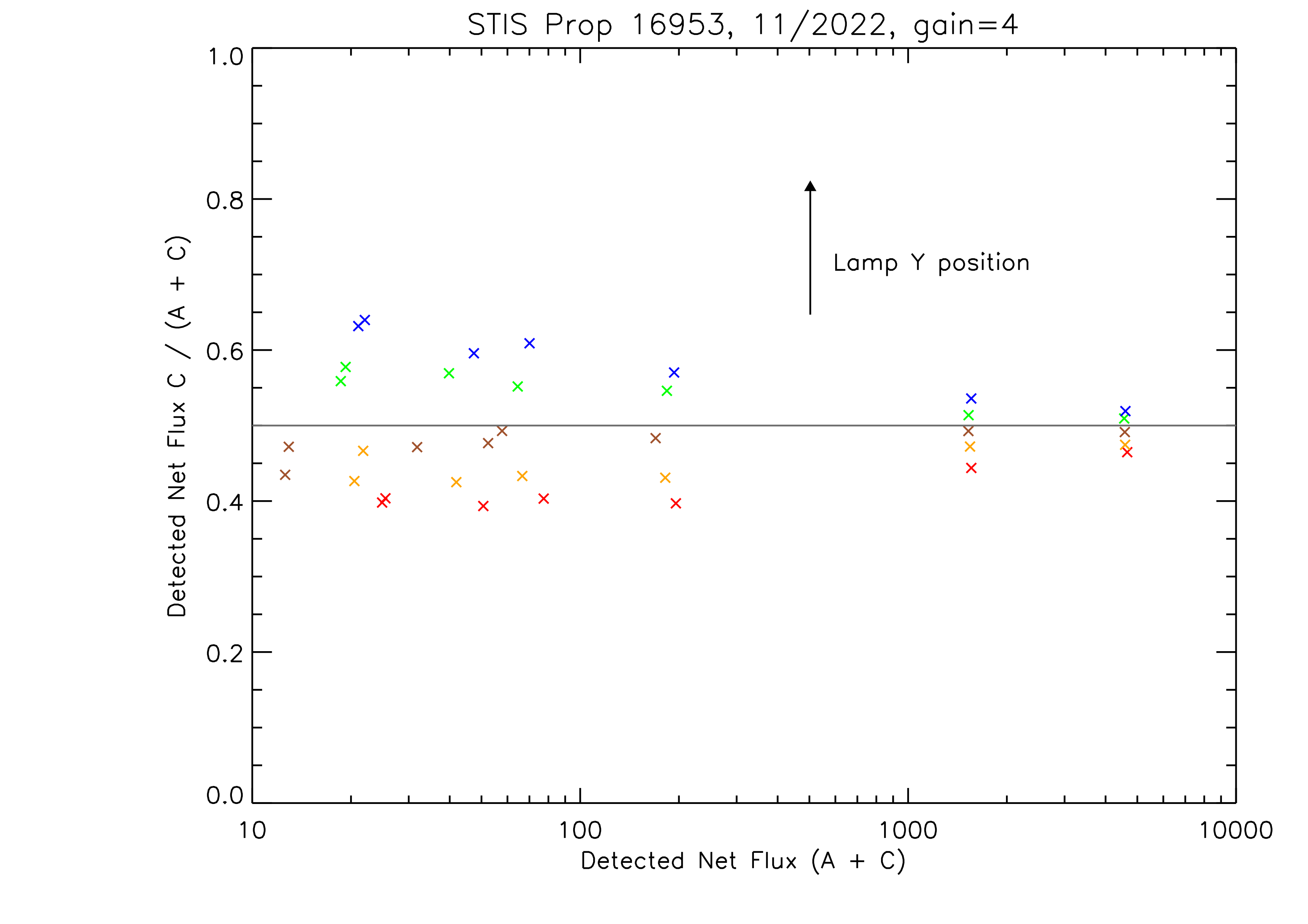}
  \includegraphics[width=120mm]{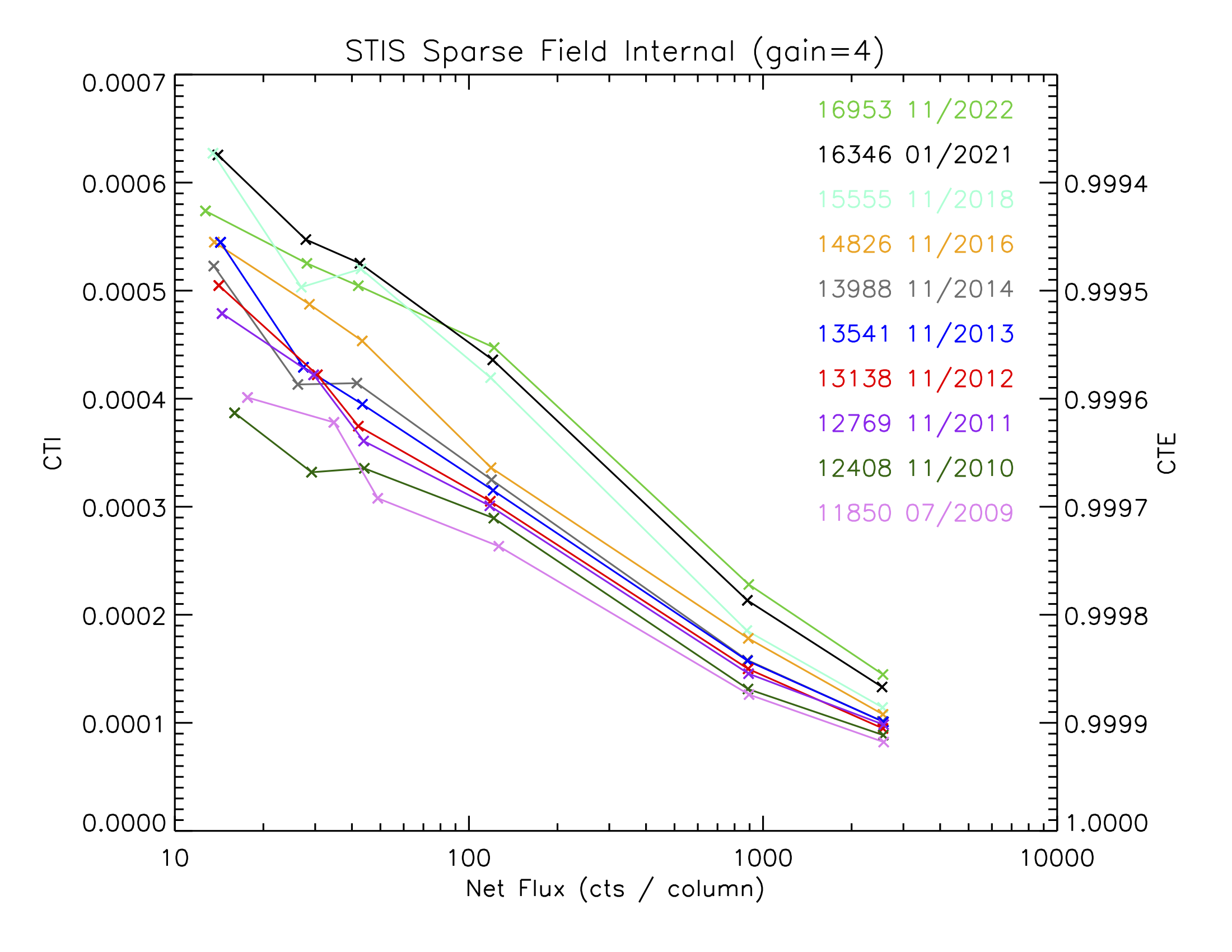}
    \caption{({\it top}) Flux ratio C/(A+C) vs. flux, for different y-positions (color-coded).
The effects of CTI are most significant at low flux levels.
({\it bottom}) Gain=4 CTI (as derived from the upper plot) vs. flux, for different epochs (color-coded; Cycle 30 values are in green).
The CTI appears to be increasing roughly linearly with time.  
These estimates are somewhat rough, but are useful for gauging relative CTI levels and their effects on science observations.}
    \label{fig:csparse}
\end{figure}

\noindent
{\bf Results:}
The roughly linear trend in the increase of the CTI seen in previous cycles continued in Cycle 30 (Fig.~\ref{fig:csparse}); the CTI effects are most significant at low flux levels.
Future analyses will better model the source flux by including the effects of spurious charge; further updates may be needed for some of the CTE coefficients (e.g., \href{https://www.stsci.edu/files/live/sites/www/files/home/hst/instrumentation/stis/documentation/instrument-science-reports/_documents/2022_07.pdf}{STIS ISR 2022-07}).

%%% -~-~-~-~-~-~-~-~-~- Slit Wheel -~-~-~-~-~-~-~-~-~- %%%

\clearpage
%\vspace{-0.3cm}
\ssectionstar{STIS Slit Wheel Repeatability (16958; A. Jones)}\label{sec:sec_slit}

{\bf Objectives:}
The goal of this program is to check the repeatability of the positioning of the STIS slit wheel, which contains the apertures and slits for spectroscopy and the clear, filtered, and coronagraphic apertures for imaging. 

\noindent
{\bf Observations:}
Once per year, a sequence of lamp spectra with the G230MB/2697 setting is observed through the three narrowest long slits (52X0.2, 52X0.1, 52X0.05) -- switching between the slits in various combinations. 

\noindent
{\bf Analysis:}
Each lamp exposure is cross correlated with the first exposure in its slit series to compute SHIFTA1 and SHIFTA2 zero point offsets in the dispersion and spatial directions, respectively. 
For each slit, ''peak-to-peak'' differences between the maximum and minimum SHIFTA1 and SHIFTA2 values and RMS values are computed.  

\begin{figure}[!hb]
  \centering
  \includegraphics[width=130mm]{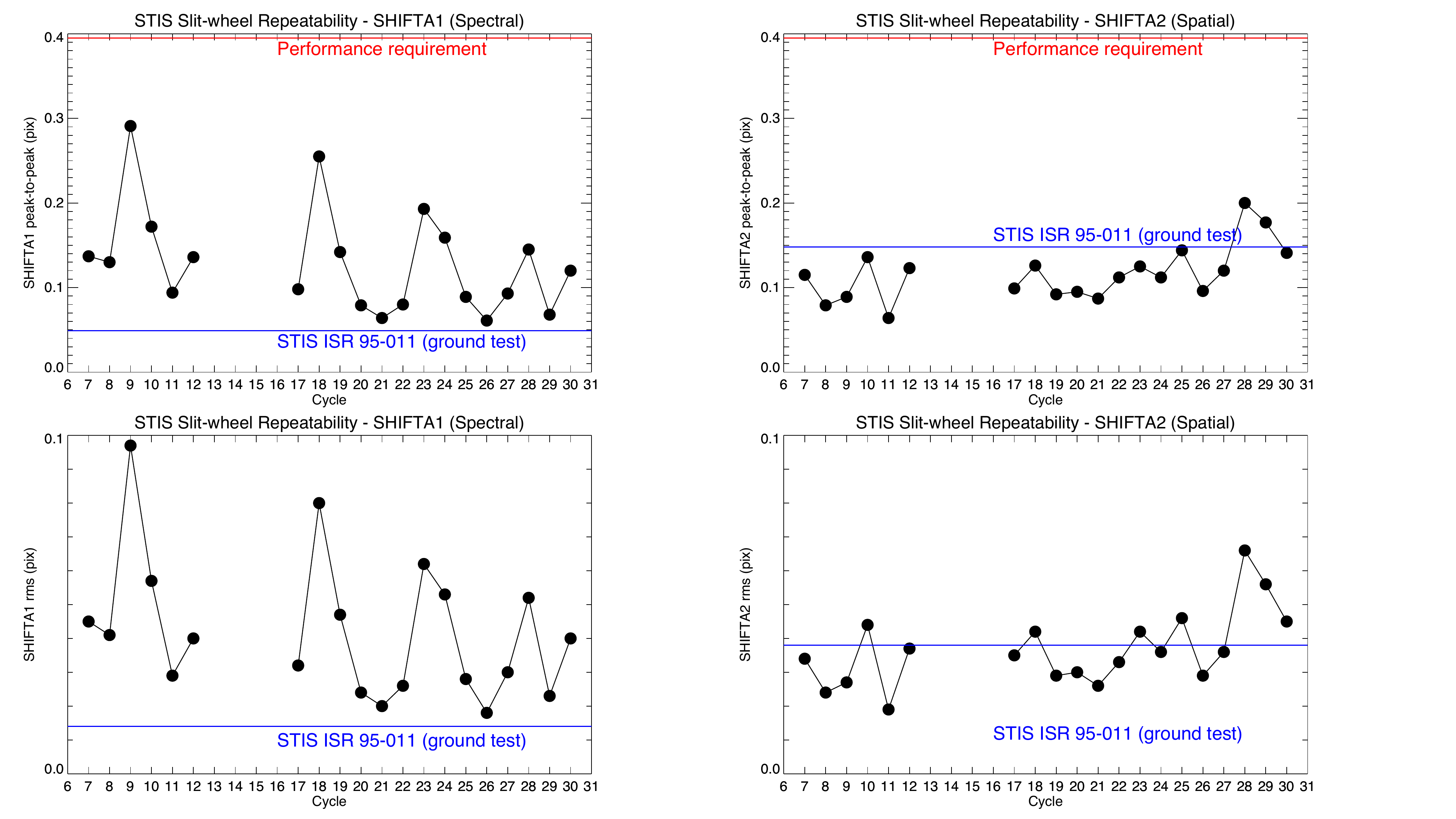}
    \caption{Measured peak-to-peak and RMS deviations of SHIFTA1 (spectral; left) and SHIFTA2 (spatial; right), compared to pre-launch test values (blue) and the desired limits (red).}
    \label{fig:slitwh}
\end{figure}

\noindent
{\bf Results:}
Slit-wheel positioning is typically repeatable to very high precision: $\pm$2.5 and $\pm$7.5 mas in the spectral/dispersion and spatial directions, respectively.
Figure~\ref{fig:slitwh} (left) shows the average peak-to-peak (top) and RMS (bottom) SHIFTA1 values for Cycles 7-30, compared to the nominal STIS CCD peak-to-peak performance requirements of 20 mas (0.395 CCD pixels; in red) and the pre-flight ground test specifications (in blue). 
In Cycles 29-30, the average peak-to-peak differences have been 0.068-0.12 pixels (3.4-6.1 mas) -- more than 0.25 pixels below the stated CCD performance requirements. 
Figure~\ref{fig:slitwh} (right) shows the peak-to-peak averages and RMS values for SHIFTA2.
In Cycles 29-30, the peak-to-peak values have been 0.141-0.177 pixels (7.1-9.0 mas) -- roughly 0.2 pix better than the requirements. 
For both SHIFTA1 and SHIFTA2, the RMS values are currently below 0.05 pix (2.5 mas).
While the peak-to-peak measurements do not account for thermal focus drifts during an orbit (which can create systematic trends), inclusion of those effects would not violate the accuracy specifications.

%%% -~-~-~-~-~-~-~-~-~- Dispersion Solutions -~-~-~-~-~-~-~-~-~- %%%

\clearpage
\ssectionstar{CCD Spectroscopic Dispersion Solution Monitor (16954; D. Welty)}\label{sec:sec_cdisp}
\vspace{-0.4cm}
{\bf ~~~~~and}
\vspace{-0.8cm}
\ssectionstar{MAMA Spectroscopic Dispersion Solution Monitor (16959; D. Welty)}\label{sec:sec_mdisp}

{\bf Objectives:}
These programs monitor the wavelength calibration / dispersion solutions and the brightness of the internal Pt/Cr-Ne lamps, for a selection of STIS CCD and MAMA settings.

\noindent
{\bf Observations:}
For the CCD, deep wavecals are obtained for representative wavelength settings with all six gratings (G230LB, G230MB, G430L, G430M, G750L, G750M), during three visits executed within 48 hr, generally near the beginning of the cycle.
These deep wavecals are generally much longer than the default wavecals obtained with most science exposures, to enable measurement of many of the weaker lamp lines that are needed to assess the behavior of the wavelength solution over the full range of each setting.
All CCD observations are currently obtained with the HITM1 Pt/Cr-Ne lamp, through the 52x0.1 aperture (which maps to 2 pixels at the CCD) and at a lamp current of 10 mA.
Since Cycle 26, the longer exposures have used CR-SPLIT=2 to enable rejection of cosmic rays. \\
For the MAMAs, deep wavecals are obtained at primary and secondary central wavelengths for all eight MAMA gratings (G140L, G140M, G230L, G230M, E140H, E140M, E230H, E230M), during seven visits executed within 96 hr, generally near the beginning of the cycle.
Most of the MAMA settings use the LINE Pt/Cr-Ne lamp, but two of the shortest wavelength settings use the somewhat brighter HITM2 Pt/Cr-Ne lamp.
Several apertures are used, and all but one setting use a lamp current of 10 mA.

\noindent
{\bf Analysis:}
The wavecal spectra are treated as science exposures, but with no background corrections, flux calibration, or Doppler corrections.  
The wavelength scale is assigned via the standard dispersion relations, with the zero point determined from the data themselves.  
The wavelengths of discernible emission lines are measured, then compared with a list of laboratory wavelengths to determine the mean offset between the measured and lab wavelengths, the scatter in the residuals, and any systematic trends in the residuals.  
Some screening is performed on the detected emission lines to try to exclude spurious and/or blended features from the statistics.  
A measure of the strength of each detected line is computed, and those line strengths are compared with previous measurements to track the fading of the lamps as a function of wavelength. 
See \href{https://www.stsci.edu/files/live/sites/www/files/home/hst/instrumentation/stis/documentation/instrument-science-reports/_documents/2018_04.pdf}{STIS ISR 2018-04} for more details.

\noindent
{\bf Results:}
All the CCD and MAMA data for Cycle 30 were obtained in 2022 Nov.  
The exposure times remain adequate (given the fading of the lamps) for measurement of an adequate set of lines; and all the first-order CCD and MAMA settings have been analyzed (at the nominal central y-location only).
All first-order settings remain within the desired tolerances for the mean wavelength offsets and standard deviations (both are$<$0.3 pix at the center of the detectors), but several CCD settings have in recent years exhibited increasingly large zero-point offsets ($>$0.5 pix) at the E1 pseudo-apertures (y $\sim$ 900).
The behavior of the offsets at E1 (and at other locations along the long slits) is suggestive of a systematic rotation of the CCD format -- as indicated also by the observed rotation of the spectral traces (\href{https://www.stsci.edu/files/live/sites/www/files/home/hst/instrumentation/stis/documentation/instrument-science-reports/_documents/200703.pdf}{STIS ISR 2007-03}) and of the dust motes seen in the CCD imaging flats (\href{https://www.stsci.edu/files/live/sites/www/files/home/hst/instrumentation/stis/documentation/instrument-science-reports/_documents/2022_01.pdf}{STIS ISR 2022-01}).
The measured line strengths indicate that the Pt/Cr-Ne lamps continue to fade, consistent with recent trends (e.g., \href{https://www.stsci.edu/files/live/sites/www/files/home/hst/instrumentation/stis/documentation/instrument-science-reports/_documents/2017_04.pdf}{STIS ISR 2017-04} and \href{https://www.stsci.edu/files/live/sites/www/files/home/hst/instrumentation/stis/documentation/instrument-science-reports/_documents/2018_04.pdf}{STIS ISR 2018-04}; see also Fig.~\ref{fig:ratios3} in Appendix A).
We note that the echelle mode data have not been analyzed since Cycle 21 (though see \href{https://ui.adsabs.harvard.edu/abs/2022AJ....163...78A/abstract}{Ayres 2022}).

\clearpage

\begin{figure}[!h]
  \centering
  \includegraphics[width=150mm]{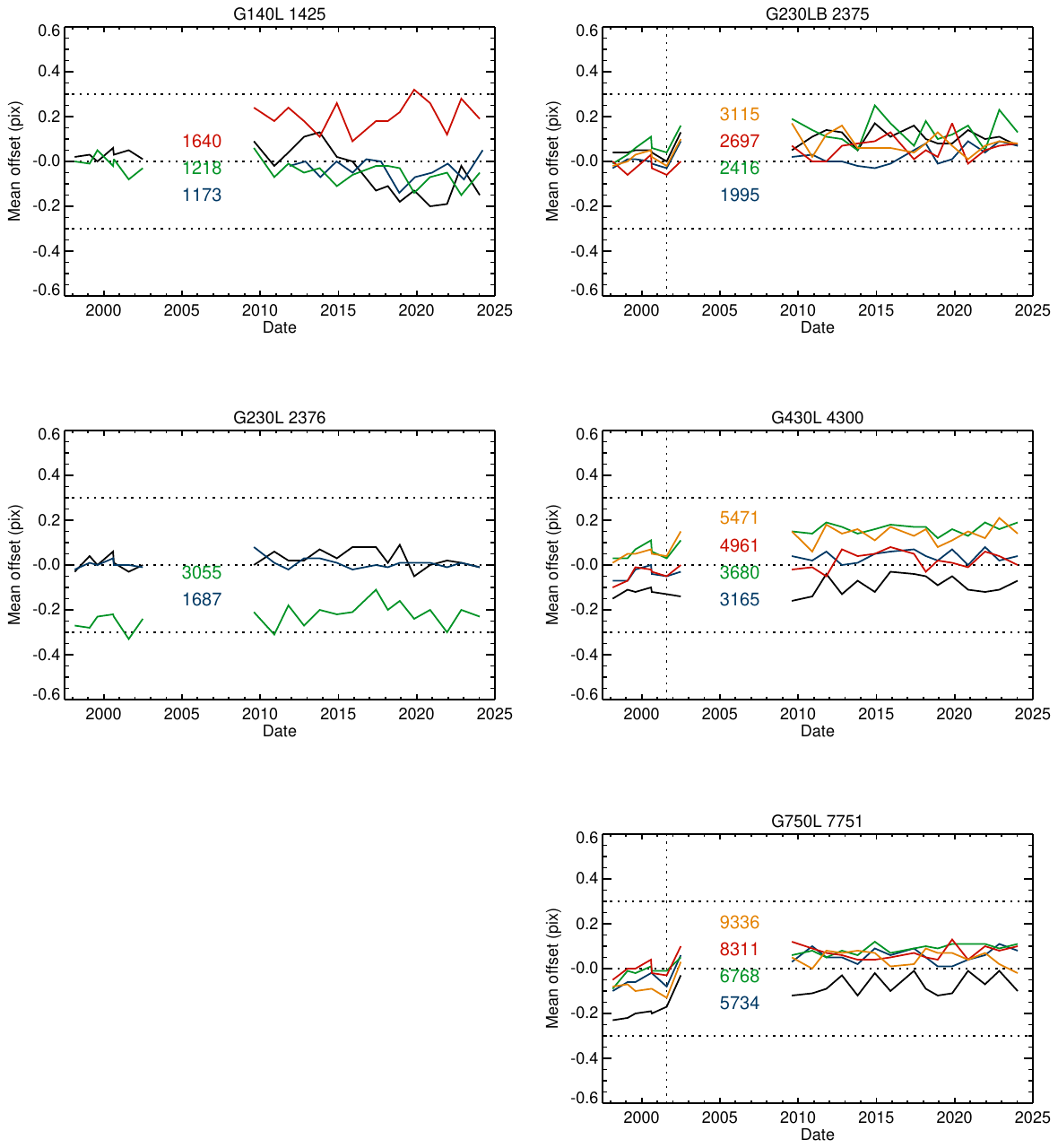}
    \caption{Mean wavelength zero point offsets (in pixels) for CCD (right) and MAMA (left) settings.
             For each panel, values for the L-mode (given in the title) are in black, and for the M-modes (noted in the body of the panel) are in color.
             The mean offsets, for spectra at the nominal central location on the detectors, all remain below 0.3 pix.}
    \label{fig:dispmon}
\end{figure}

%%% -~-~-~-~-~-~-~-~-~- Full Field Sensitivity -~-~-~-~-~-~-~-~-~- %%%

\clearpage
%\vspace{-0.3cm}
\ssectionstar{CCD Full Field Sensitivity Monitor (16955; S. Lockwood / E. Rickman)}\label{sec:sec_cffsens}
\vspace{-0.4cm}
{\bf ~~~~~and}
\vspace{-0.8cm}
\ssectionstar{MAMA Full Field Sensitivity Monitor (16956; S. Lockwood / E. Rickman)}\label{sec:sec_mffsens}

{\bf Objectives:}
These programs monitor the sensitivity and the astrometric and PSF stability over the full field of each detector, via observations of standard star fields.

\begin{figure}[!h]
  \centering
  \includegraphics[width=75mm]{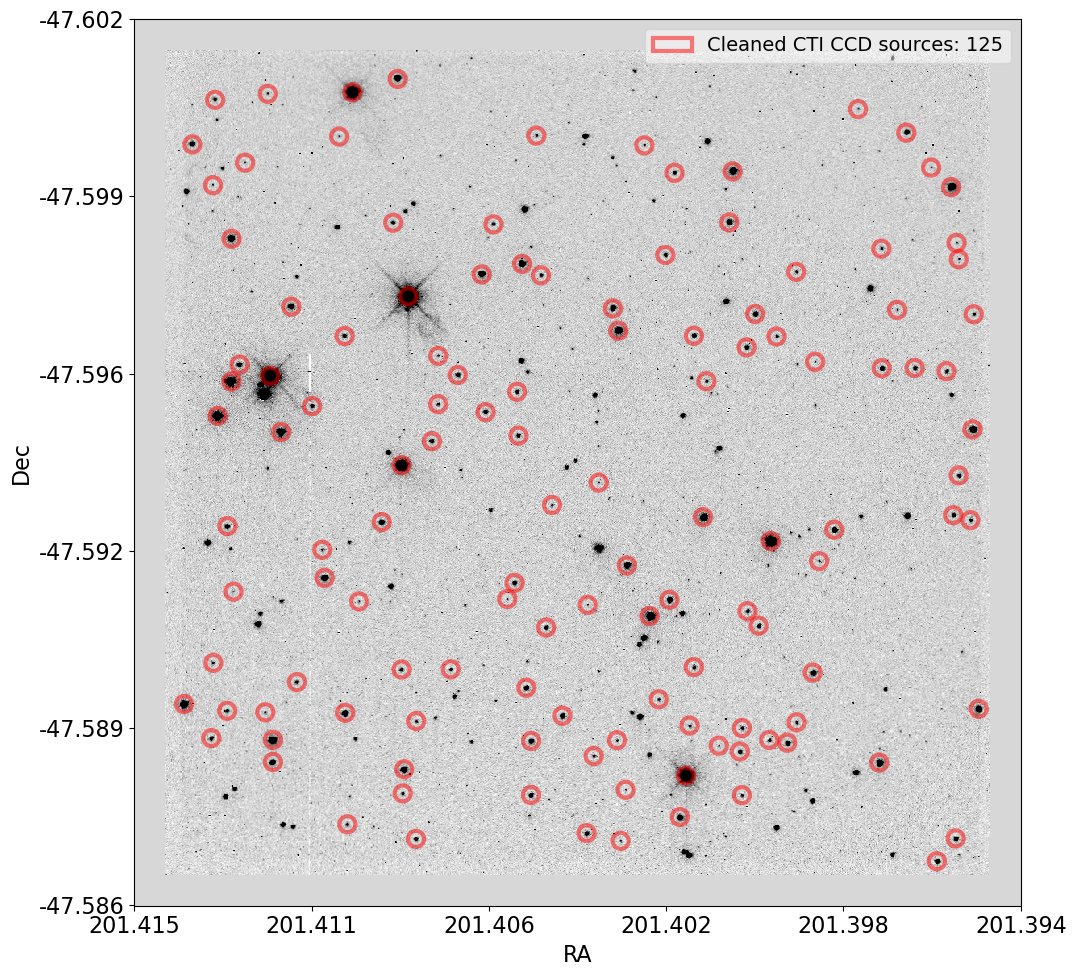}
  \includegraphics[width=75mm]{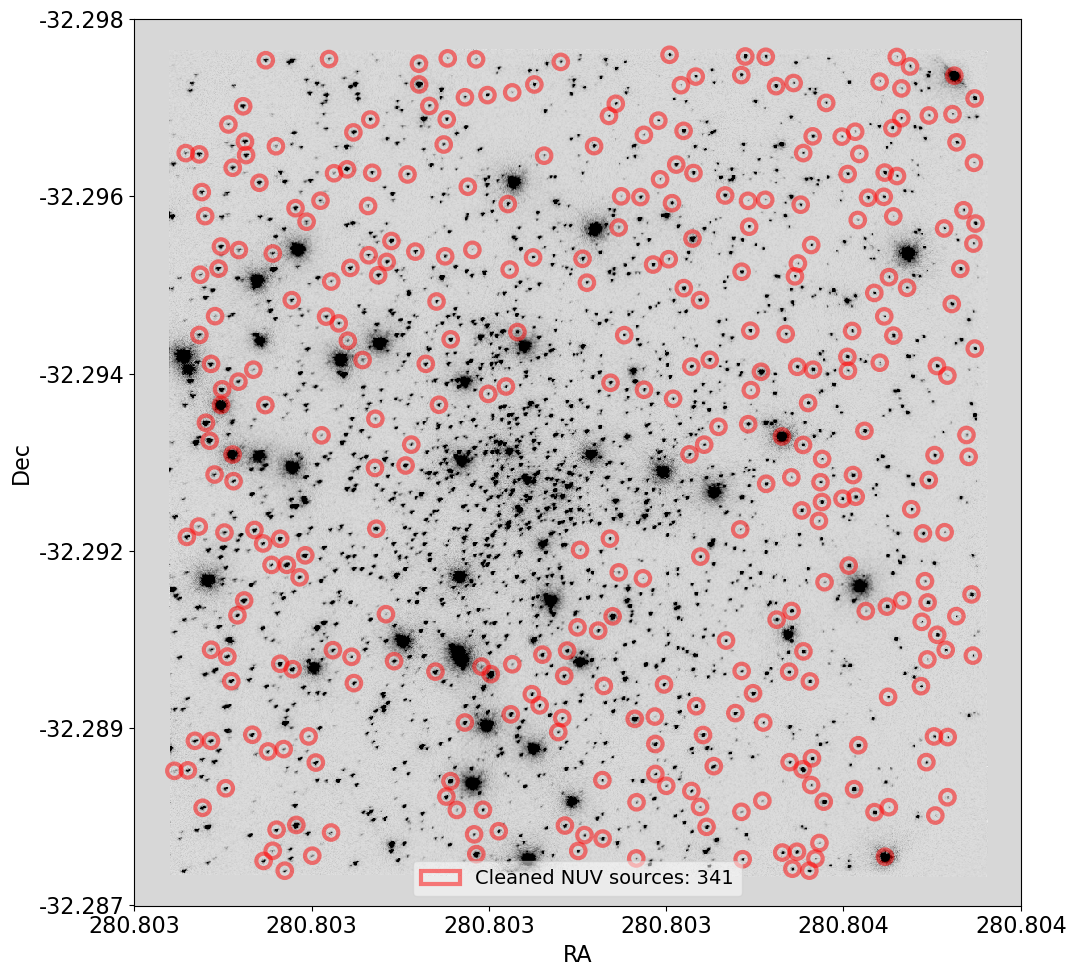}
    \caption{({\it left}) CTI-corrected CCD image of omega Cen, with 125 isolated stars used for photometric monitoring.  ({\it right}) NUV-MAMA image of NGC 6681, with 341 isolated stars used for photometric monitoring.}
    \label{fig:ffsens}
\end{figure}

\noindent
{\bf Observations:}
For the CCD, standard fields in omega Cen (Fig.~\ref{fig:ffsens}) are observed through the 50CCD aperture, at a standard ORIENT of 310 degrees (as for previous cycles), in a single orbit, once per year.
Exposures are taken at several positions around the nominal center, for 10s and 60s, and for amps A, C, and D; all are at gain=4.
For the MAMAs, a standard field in NGC6681 (Fig.~\ref{fig:ffsens}) is observed through the F25SRF2 (x4), F25QTZ, and F25CN182 apertures for the NUV and through the 25MAMA (x4), F25QTZ, and F25SRF2 apertures for the FUV -- in a single 3-orbit visit, at the same ORIENT (260-266 degrees) as previous observations.

\noindent
{\bf Analysis:}
Aperure photometry is performed for all adequately isolated stars in the field of view, which also allows monitoring of the PSF.
The geometric distortion map can be checked by computing the polynomial transformation required to map the HST reference coordinate system to the coordinate system of the STIS detectors.
The MAMA observations can also be used to look for contamination, throughput changes, or formation of defects in the photocathode and window that might be missed by spectroscopic monitoring or difficult to interpret in flat-fielding.

\noindent
{\bf Results:}
\href{https://www.stsci.edu/files/live/sites/www/files/home/hst/instrumentation/stis/documentation/instrument-science-reports/_documents/2022_02.pdf}{STIS ISR 2022-02} presents the results from a uniform analysis of the CCD and MAMA full-field sensitivity data through Cycle 29.
While the residual TDS trends for most of the apertures are consistent with the desired 1\% STIS flux calibration accuracy, the imaging sensitivity appears to be declining more slowly than would be expected from the spectroscopic TDS analyses.
 
%%% -~-~-~-~-~-~-~-~-~- Spectro Sensitivity -~-~-~-~-~-~-~-~-~- %%%

\clearpage
%\vspace{-0.3cm}
\ssectionstar{CCD Spectroscopic Sensitivity Monitor (16957; S. Hernandez / D. Stapleton)}\label{sec:sec_cspsens}
\vspace{-0.4cm}
{\bf ~~~~~and}
\vspace{-0.8cm}
\ssectionstar{MAMA Spectroscopic Sensitivity and Focus (16960; S. Hernandez / D. Stapleton)}\label{sec:sec_mspsens}

{\bf Objectives:}
Data obtained for these programs are used to monitor the spectroscopic sensitivity of the STIS CCD and MAMA detectors, in selected low-, medium-, and high-resolution settings, for flux calibration and exposure time calculations.
The CCD data can also be used to monitor the effects of charge transfer inefficiency.
Additional imaging observations are used to assess the focus during the MAMA visits.

\noindent
{\bf CCD Observations:}
All three low-resolution (``L'') modes are monitored in 1-orbit visits, three times per year.
Five representative medium-resolution (``M'') mode settings are monitored in a single annual 2-orbit visit.
All observations are of the high-latitude standard star AGK+81$^\circ$266, through the 52x2 aperture; all of the settings are observed both at the nominal central position on the CCD and at the E1 pseudo-aperture (y $\sim$ 900).
The first Cycle 30 L-mode visit failed twice due to guide star issues, resulting in a change in the cadence of the three L-mode observations -- both within Cycle 30 and relative to the last visit of Cycle 29.
The other three visits executed successfully. \\
{\bf MAMA Observations:}
G140L and G230L observations of GRW+70$^\circ$5824 are obtained in 1-orbit visits three times per year.
G140M and G230M observations of AGK+81$^\circ$266, at several wavelengths, are obtained in a 1-orbit visit once per year.
Echelle observations of BD+28$^\circ$4211, at several wavelengths, are obtained in 2-orbit visits four times per year.
For the L- and M-mode visits, the initial focus is determined via narrow-band (F28X50OII) images, supplemented by CCD/G230LB spectra for the L modes.
Since Cycle 27, parallel observations with WFC3 have been used to monitor the focus on orbital time scales during the echelle visits, as focus variations can affect the throughput (see \href{https://www.stsci.edu/files/live/sites/www/files/home/hst/instrumentation/stis/documentation/instrument-science-reports/_documents/2017_01.pdf}{STIS ISR 2017-01}).
All of the Cycle 30 visits executed as planned.

\noindent
{\bf Analysis:}
The analysis of the time-dependent sensitivity (TDS) of the STIS detectors is described in \href{https://www.stsci.edu/files/live/sites/www/files/home/hst/instrumentation/stis/documentation/instrument-science-reports/_documents/2014_02.pdf}{STIS ISR 2014-02} and \href{https://www.stsci.edu/files/live/sites/www/files/home/hst/instrumentation/stis/documentation/instrument-science-reports/_documents/2017_06.pdf}{STIS ISR 2017-06}.
For each observation, the net counts (corrected for known variable effects -- e.g., CTE, temperature, and ``red halo'' for the CCD) are determined for a fixed set of wavelength bins, then normalized by the corresponding values from the initial execution of the program in 1997.
For each wavelength bin, the normalized counts vs. time are fitted with a segmented line model.
The fits are parameterized by the number of line segments, the breakpoints defining the beginnings of the segments, and the slopes of the line segments.

\begin{figure}[!h]
  \centering
  \includegraphics[width=170mm]{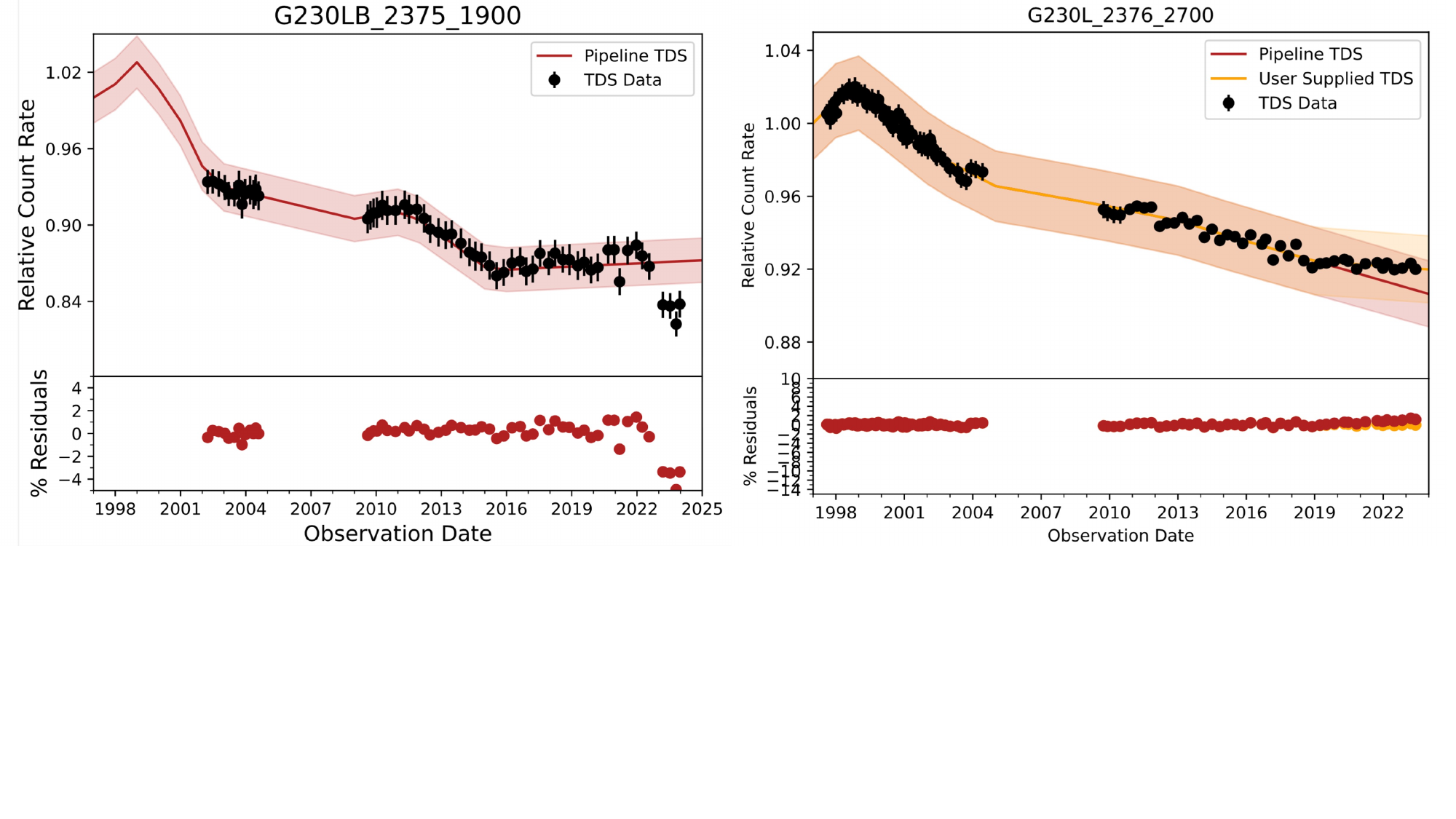}
  \vspace{-3.5cm}
    \caption{({\it left}) Observed trends in net count rates, relative to 1997, for G230LB between 1900 and 2100\AA\ (through Cycle 30). The solid line is the segmented line fit to the data (pre-Cycle 30); the shaded region shows $\pm$2\% about that fit.  Note the sharp drop in this spectral bin for the most recent observations.
({\it right}) Observed trends in net count rates, relative to 1997, for G230L between 2700 and 2800 \AA. The solid red line is the segmented line fit to the data (pre-Cycle 30); the shaded region shows $\pm$2\% about that fit.  The solid orange line and corresponding shaded region represent the new fit, with a new, flatter segment starting at $\sim$2018.}
    \label{fig:cmtds}
\end{figure}

\noindent
{\bf Results:}
In recent cycles, the monitored wavelength settings have generally exhibited either roughly constant sensitivities or slow declines. \\
{\bf CCD:} 
For Cycle 30, the monitored settings for G430L, G430M, G750L, and G750M all appear to be consistent with the recent trends -- both at the nominal central location and at E1.
For G230LB, however, the Cycle 30 data have indicated a sudden dip in the sensitivity in the 1900-2100 \AA\ bins (Fig.~\ref{fig:cmtds}, left-hand panel); that dip appears to be corroborated by the observations of G230MB/1995, as well as by nearly contemporaneous observations at those wavelengths obtained with G230L and the NUV MAMA.
[Note: A special calibration program (PID 17584) was instituted during Cycle 31 to obtain more frequent observations of the NUV sensitivity.] \\
{\bf MAMAs:}
For Cycle 30, the monitored FUV and NUV settings generally appear to be consistent with the recent trends, but there are a few exceptions: 
\begin{itemize}
\item{G140L -- a possible change in the trend (across all wavelengths) -- this will be monitored}
\item{G230L -- a sudden dip in sensitivity at 1900-2100 \AA\ (as seen also for G230LB)}
\item{G230L -- a flattening of the trend for $\lambda$$>$2400 \AA\ (Fig.~\ref{fig:cmtds}, right-hand panel)}
\end{itemize}
The G140M and G230M settings are consistent with previous trends; G140M/1173 continues to lie slightly above the adopted pipeline trend (investigations are on-going).
The echelle settings (E140H, E140M, E230H, E230M) also are consistent with previous trends, though some have larger scatter ($\sim$8\% for E230M) and/or misalignment vs. the adopted pipeline TDS. 
We note, however, that the TDS trends are impacted by the accuracy of the blaze shift corrections, which are being worked on.

\noindent
{\bf Files:}
In 2023~Jul, a new TDSTAB was generated to add a breakpoint for the NUV-MAMA settings longward of 2400 \AA; corresponding synphot files were delivered for the ETC.
Other routine synphot updates (both CCD and MAMA) were delivered in 2023~Oct.
[Note: Additional changes to the NUV sensitivity near 1900-2100 \AA, based on analyses of data from special program 17584, were delivered in 2024 July; see the \href{https://www.stsci.edu/contents/news/stis-stans/march-2024-stan#article3}{2024 March STAN}.] 

%%% -~-~-~-~-~-~-~-~-~- MAMA Darks -~-~-~-~-~-~-~-~-~- %%%

\clearpage
%\vspace{-0.3cm}
\ssectionstar{FUV MAMA Dark Monitor (16961; S. Lockwood)}\label{sec:sec_fmdark}
\vspace{-0.4cm}
{\bf ~~~~~and}
\vspace{-0.8cm}
\ssectionstar{NUV MAMA Dark Monitor (16962; A. Jones)}\label{sec:sec_nmdark}

{\bf Objectives:}
These programs monitor the time and temperature variations of the dark current of the two MAMA detectors.

\noindent
{\bf Observations:} \\
{\bf FUV:} sets of six 1300s dark exposures (with shutter closed), in six consecutive orbits, every six weeks.
That cadence enables tracking of the FUV glow (which increases with increasing detector time on, especially after orbit 2) over the full range of allowed STIS visit lengths. \\
{\bf NUV:} sets of two 1300s dark exposures, in two orbits spaced by 3-4 orbits, every two weeks.
That cadence allows both short- and longer-term variations to be discerned -- which can be particularly valuable after recovery from a safing (which can lead to elevated dark rates for a few weeks thereafter).
The exposure lengths are designed to achieve a 1\% statistical error on the total counts in each exposure.

\noindent
{\bf Analysis:} For both FUV and NUV, the temperature-related variations (typically roughly quadratic) are determined and removed -- so that the remaining temporal variations may be assessed. 
The analysis of the FUV darks is discussed in \href{https://www.stsci.edu/files/live/sites/www/files/home/hst/instrumentation/stis/documentation/instrument-science-reports/_documents/2015_07.pdf}{STIS ISR 2015-07}.
For the NUV, the temperature-corrected dark rates are fitted with exponential declines following significant safing events and smooth variations thereafter.
These analyses provide the dark rate scaling values in the TDCTAB reference file, which is used for the pipeline processing of science data.

\begin{figure}[!h]
  \centering
  \includegraphics[width=120mm]{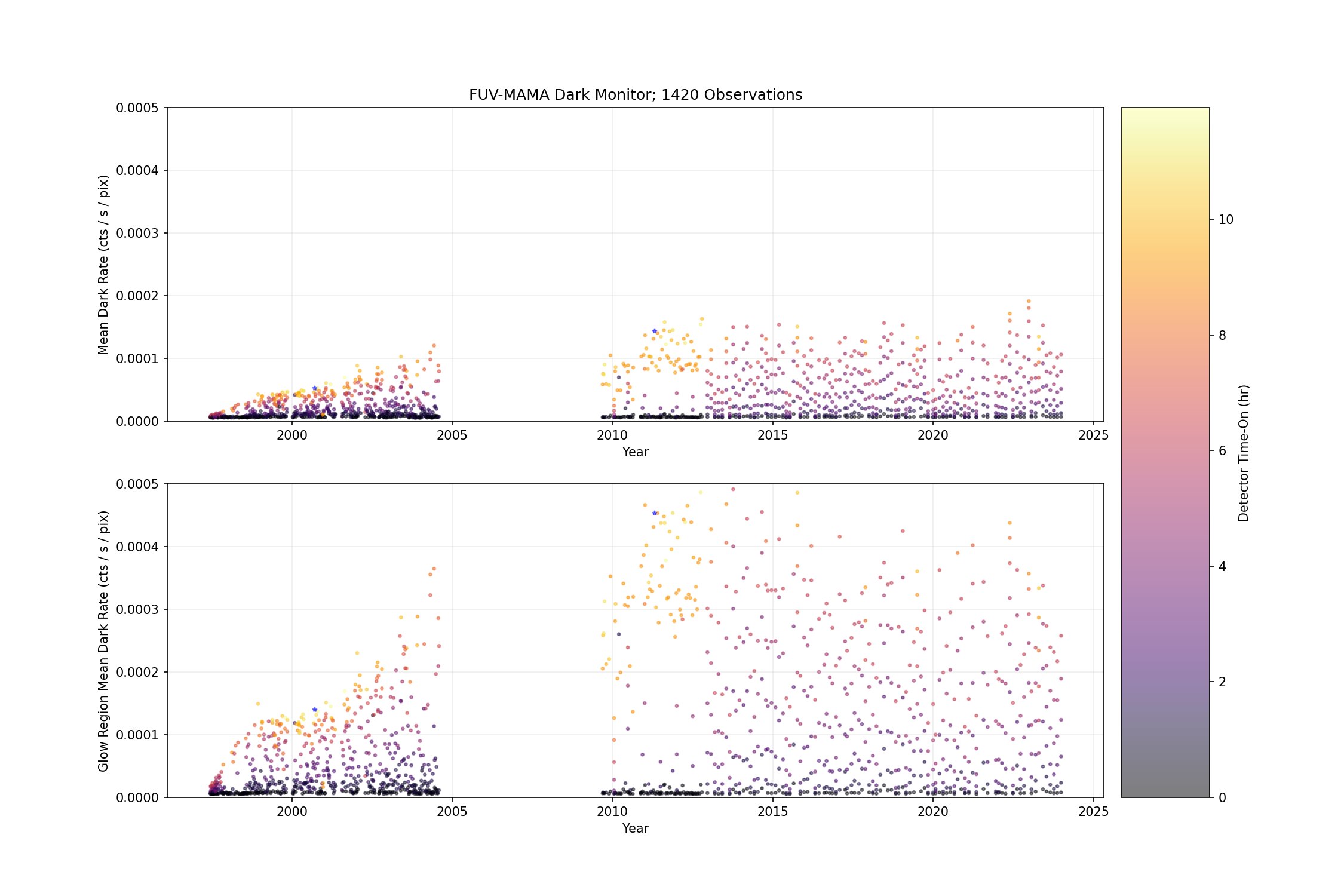}
    \caption{Dark count rates for the FUV MAMA, color coded by detector time on.  The top plot is for the full detector; the bottom plot is for the ''glow'' region, located in the upper left-hand part of the detector.  The rates in the glow region increase dramatically after orbit 2 of each 6-orbit visit.}
    \label{fig:fmdark}
\end{figure}

\begin{figure}[!h]
  \centering
  \includegraphics[width=120mm]{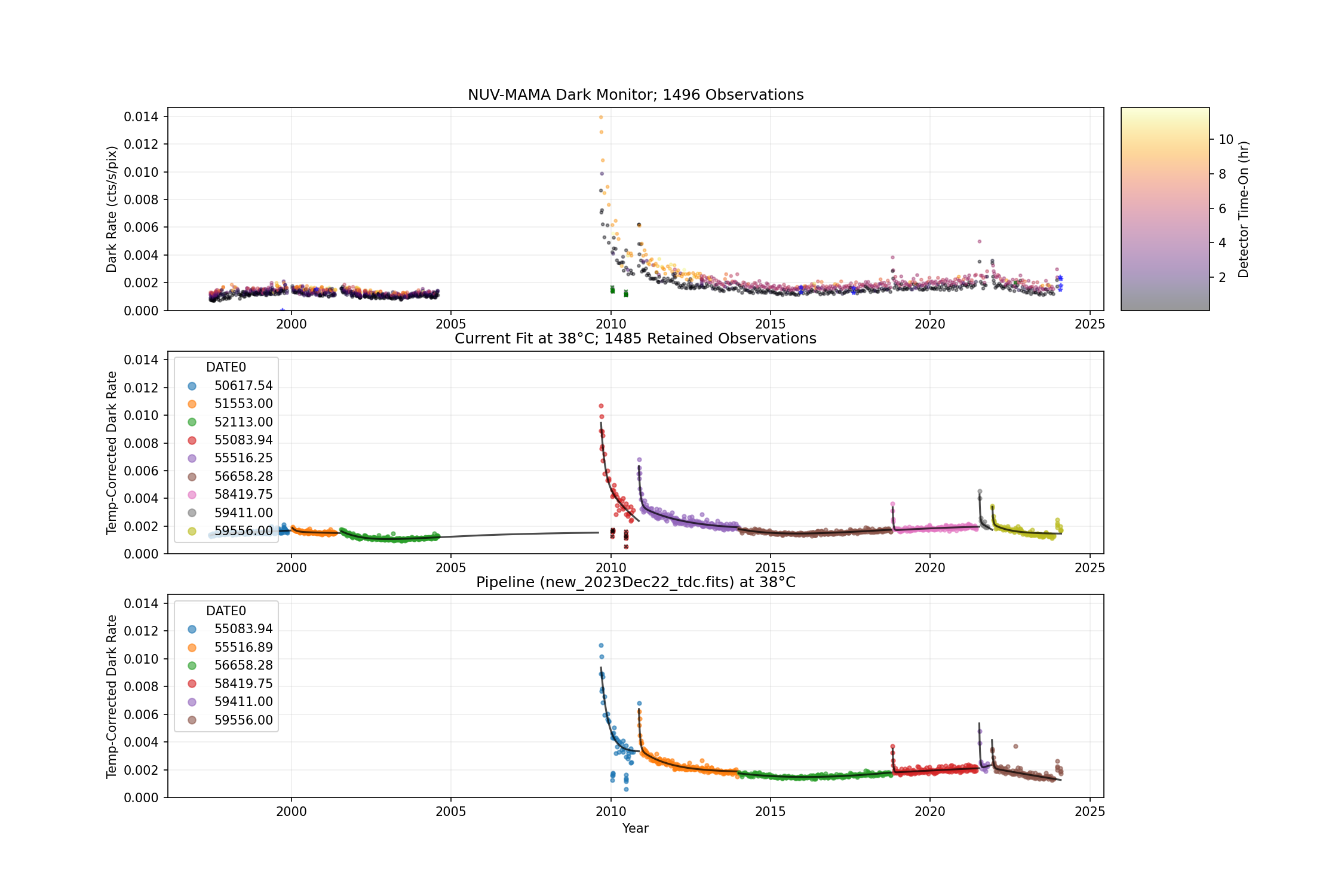}
    \caption{Dark count rates for the NUV MAMA.  Discontinuities, marked by spikes in the dark rates followed by declines to more ``normal'' levels, correspond to recoveries from HST safing events in which the STIS detectors were off for an extended period of time.  The different colors in the lower two plots denote the different intervals over which the temperature-corrected rates are fitted.}
    \label{fig:nmdark}
\end{figure}

\noindent
{\bf Results:} \\
{\bf FUV:} New, somewhat lower FUV dark rates, based on recent monitoring observations (Fig.~\ref{fig:fmdark}), were adopted for ETC 31.1 -- a base rate of 3e-5 cts/sec/pix (for 1-2 orbits after high voltage turn-on), with additional glow region contributions for increasing detector time on:  low glow = 4.5e-5 cts/sec/pix (after 2-3 orbits), medium glow = 1.2e-4 cts/sec/pix (4 orbits), and high glow = 3.7e-4 cts/sec/pix (5$^+$ orbits). \\
{\bf NUV:} The NUV dark rate continues to follow the recent general trends with temperature and detector time on (Fig.~\ref{fig:nmdark}), with a slow decline below 0.002 cts/s/pix during Cycle 30.
The three HST safing events during 2023~Aug were all very brief ($\sim$1 day), and no significant increases in the NUV dark rate were noted.
%New super-dark reference files have been delivered; updates to the TDCTAB are in progress.

%%% -~-~-~-~-~-~-~-~-~- FUV MAMA Flats -~-~-~-~-~-~-~-~-~- %%%

\clearpage
%\vspace{-0.3cm}
\ssectionstar{FUV MAMA Flat Field Monitor (16963; A. Jones)}\label{sec:sec_fmflat}

{\bf Objectives:}
This program obtains flat field observations, which are used to construct pixel-to-pixel flat fields (p-flats) for the MAMA detectors.
To conserve the calibration lamps, FUV and NUV observations are obtained in alternate cycles.

\noindent
{\bf Observations:}
The FUV observations currently use the krypton lamp, the G140M/1470 setting, and the 52x0.2 aperture; the NUV observations currently use the deuterium lamp, the G230M/2499 setting, and the 52x0.5 aperture.
Because the lamps are fading, the central wavelength and/or aperture have been adjusted (as needed) to obtain global count rates between 180000 ct/s (to enable the desired 1\% accuracy per low-resolution pixel) and 280000 ct/s (the non-linearity limit).
Eleven long (4400 s) exposures are obtained through the year, using five distinct slit offsets in order to cover the areas blocked by the occulting bars in the long slits.

\noindent
{\bf Analysis:}
Large-scale trends and any spectral lines are removed from the individual images, then the images are combined to produce a cycle-level p-flat, taking account of the differing locations affected by the occulting bars.
The cycle-level p-flat is compared with the existing pipeline p-flat (STIS TIR 2002-03\footnotemark), and, in principle, could be folded in to produce an updated p-flat.
\footnotetext{This reference may be obtained by request through the \href{https://stsci.service-now.com/hst}{HST HelpDesk}.}

\begin{figure}[!hb]
  \centering
  \includegraphics[width=120mm]{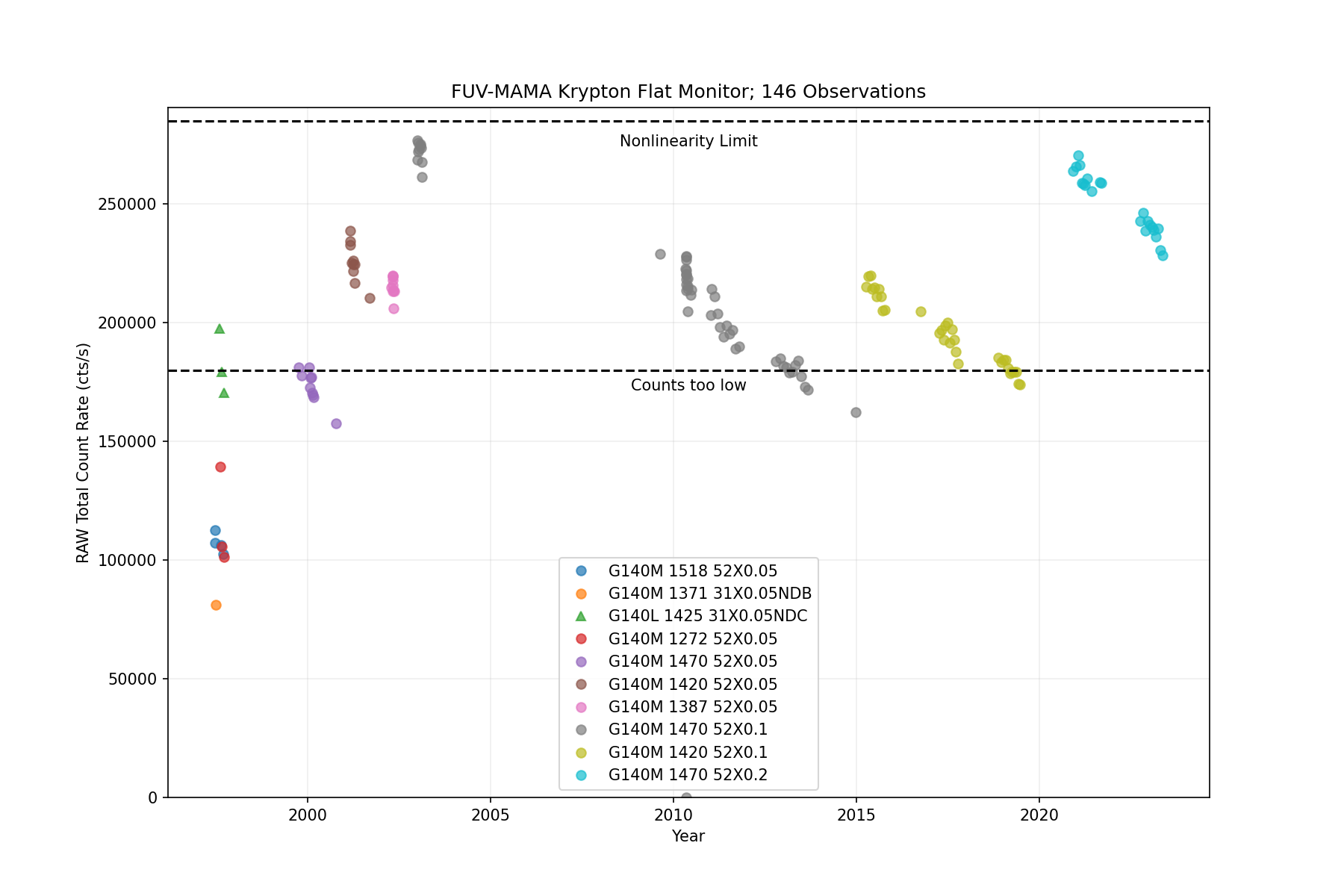}
    \caption{Global count rates for individual FUV p-flat observations.  As the Kr lamp has been fading, different wavelength settings and aperture widths have been used in order to obtain global rates between 180000 and 280000 ct/s -- to yield adequate S/N without entering the non-linear regime.}
    \label{fig:fmflat}
\end{figure}

\noindent
{\bf Results:}
For Cycle 30, the global count rates were between 250000 and 220000 ct/s, which is well within the optimal range (Fig.~\ref{fig:fmflat}).  
Comparisons with the pipeline flats suggest only very slow evolution of the flats.
The STIS team is currently updating and testing the scripts to enable creation of new p-flats, when that should become necessary.
 
%%% -~-~-~-~-~-~-~-~-~- MAMA Fold -~-~-~-~-~-~-~-~-~- %%%

\clearpage
%\vspace{-0.3cm}
\ssectionstar{MAMA Fold Distribution (16964; T. Wheeler)}\label{sec:sec_mfold}

{\bf Objectives:}
The performance of the FUV and NUV MAMA microchannel plates (MCPs) is assessed using a MAMA fold distribution analysis procedure that provides a measurement of the distribution of charge cloud sizes incident upon the anode -- giving some measure of change in the pulse-height distribution of the MCP and, therefore, MCP gain. 
The results are compared with those from previous cycles to detect trends and/or any anomalous behavior. 

\noindent
{\bf Observations:}
While globally illuminating the detector with a flat field, the valid event rate counter is monitored while various combinations of row and column folds are selected (see the \href{https://www.stsci.edu/hst/phase2-public/16964.pro}{phase II} and \href{https://www.stsci.edu/files/live/sites/www/files/home/hst/instrumentation/stis/documentation/instrument-science-reports/_documents/199802.pdf}{STIS ISR 1998-02R} for details).
This program was executed in 2 internal orbits on 2023~May~01.

\noindent
{\bf Analysis:}
The engineering telemetry data for each detector (voltages, currents, temperatures, relay positions, and status) were examined and compared with predicted values and with previous ground and on-orbit test data. 
The MAMA engineering telemetry for each detector event counter was used to construct a histogram of the number of counts for each fold, which was then compared and combined with previous test results. 
The position of the peak in the fold distribution can be measured to about 5\% accuracy from this procedure.
Post-test, a dark exposure was taken where the counters were cycled; the histogram of those data was compared with earlier results. 

\noindent
{\bf Results:}
The NUV MAMA exhibits a known high dark count rate, caused by window phosphorescence, that has been decreasing since SMOV4. 
No anomalous behavior was detected for either MAMA. 

\begin{figure}[!hb]
  \centering
  \includegraphics[width=\textwidth]{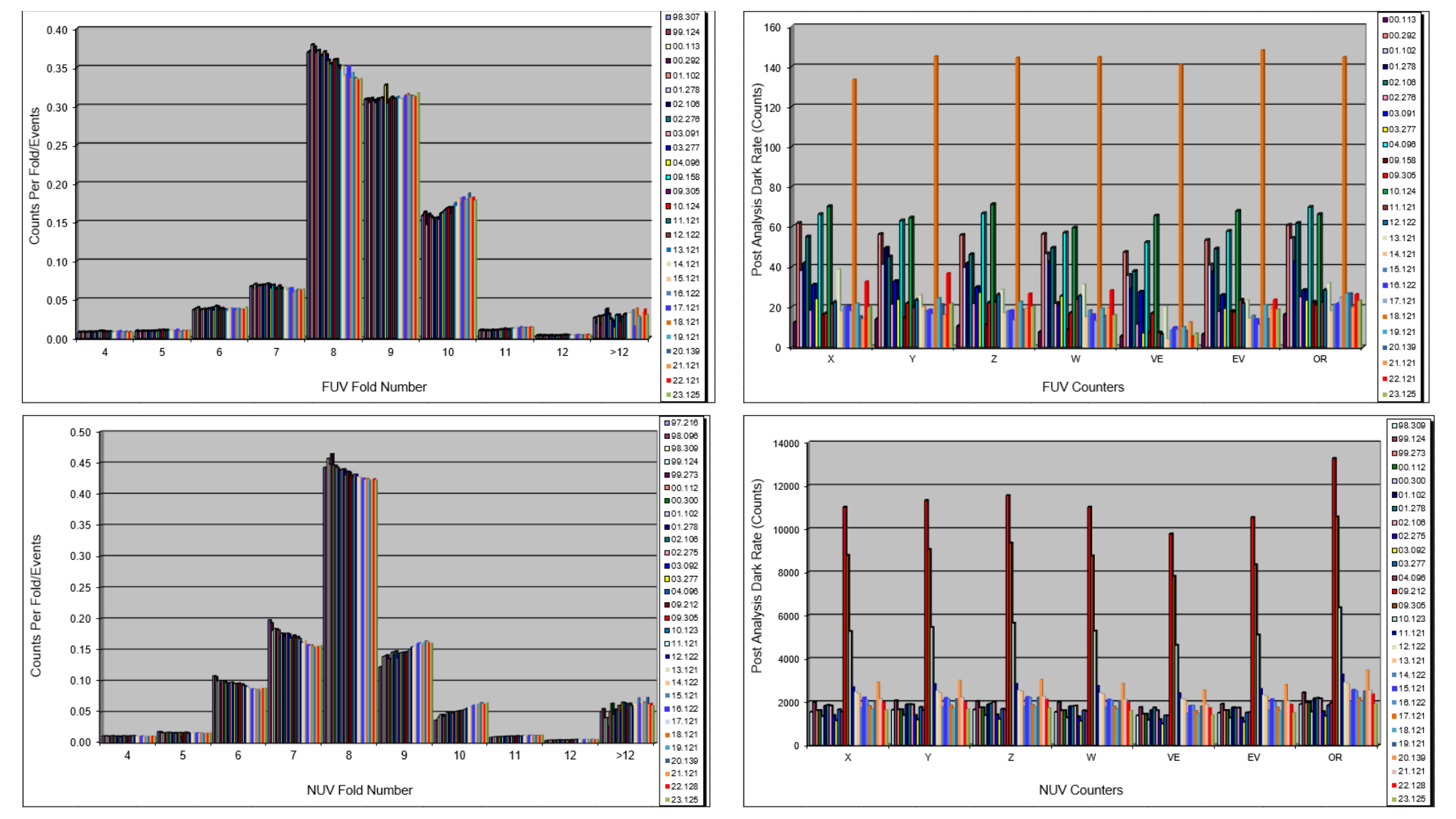}
    \caption{(top) FUV fold histogram and post-test dark count histogram.
             (bottom) NUV fold histogram and post-test dark count histogram.}
    \label{fig:fold1}
\end{figure}

%%% -~-~-~-~-~-~-~-~-~- 3 Primary WD -~-~-~-~-~-~-~-~-~- %%%

\clearpage
%\vspace{-0.3cm}
\ssectionstar{Monitoring the Three Primary WD Standard Stars (16966; R. Bohlin)}\label{sec:sec_wdmon}

{\bf Objectives:}
Regular observations of the three primary white dwarf standard stars (GD153, GD71, G191B2B) can reveal any variability in those three primary WDs, as well as in the two STIS MAMA and CCD flux monitor standards (GRW+70$^\circ$5824, AGK+81$^\circ$266). 
Each new observation adds to the statistical precision of the STIS flux calibration that is defined by those three stars. 

\noindent
{\bf Observations:}
STIS now observes the primary WD flux standards with a cadence of every two years.
Spectra are obtained for the five low dispersion modes (G140L, G230L, G230LB, G430L, G750L), all with the 52x2 slit. 
For GD153 and GD71, both MAMA and CCD spectra are obtained; for the brighter G191B2B, only the CCD modes are observed.

\noindent
{\bf Analysis:}
All spectra were obtained successfully and were reduced to flux distributions using an updated version of the Instrument Definition Team pipeline (independent of the {\sc calstis} pipeline; see, e.g., \href{https://ui.adsabs.harvard.edu/abs/2019AJ....158..211B/abstract}{Bohlin et al. 2019}). 
The average change in response for the five STIS low dispersion modes is assessed via comparisons among the three primary standards and the two STIS standards.
The uncertainty in the STIS time-dependent sensitivity correction is determined by the level of agreement between the triad of WDs and the respective monitoring standards. 

\noindent
{\bf Results:}
The changes in sensitivity seen for the three WDs are statistically consistent with those seen via the denser time coverage of the two STIS monitor stars (GRW+70$^\circ$5824, AGK+81$^\circ$266). 
When the new data are combined with the many previous observations of the three primary standards, the change in the STIS low dispersion flux calibration is much less than 1\%.

\begin{figure}[!hb]
  \centering
  \includegraphics[width=150mm]{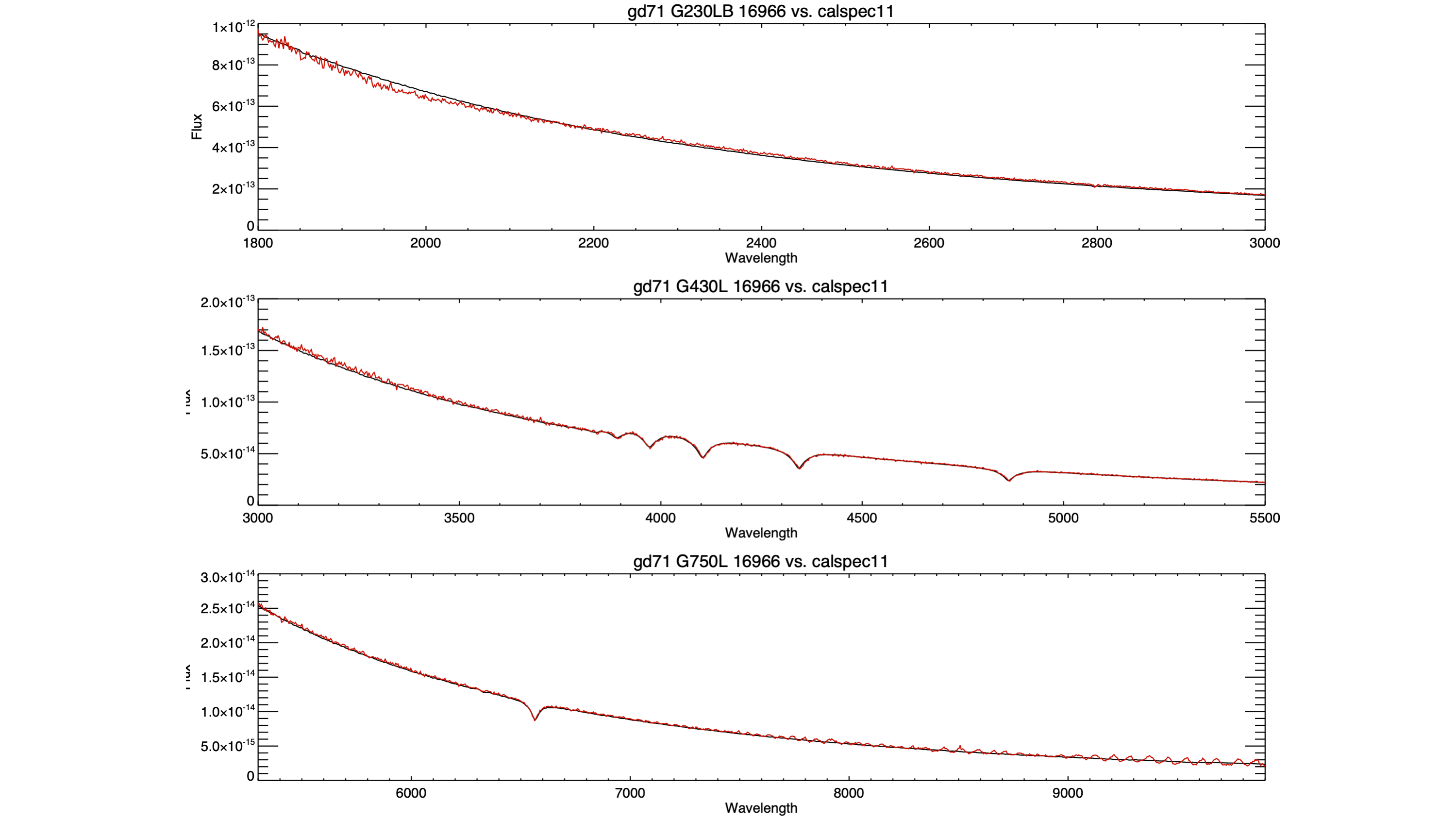}
    \caption{Pipeline-reduced STIS G230LB, G430L, and G750L spectra of GD71 from program 16966 (red) vs. CALSPECv11 spectrum (black).  Differences near 2000 \AA\ reflect a recent apparent decline in sensitivity there, which was not yet included in the pipeline calibrations (see Fig.~\ref{fig:cmtds} above).  Fringing, visible for $\lambda$ $>$ 7000 \AA, has not been removed from the G750L spectrum.}
    \label{fig:primwd}
\end{figure}

%%% DELETE THESE SECTIONS IF YOU DON'T USE THEM %%%

%%% -~-~-~-~-~- ACKNOWLEDGEMENTS, CHANGES, REFERENCES -~-~-~-~-~- %%%

\clearpage
%%% ACKNOWLEDGEMENTS %%% 
%\vspace{-0.3cm}
\ssectionstar{Acknowledgements}
%\vspace{-0.3cm}
We thank Serge Dieterich for a constructive review of the initial version of this ISR.

%%% CHANGE HISTORY %%%
\vspace{-0.3cm}
%Put instrument, year, and ISR number
\ssectionstar{Change History for STIS ISR 2025-01}\label{sec:History}
\vspace{-0.3cm}
Version 1: 21 January 2025 - Original Document %Month DD, YYYY format

%%% REFERENCES %%%
\vspace{-0.3cm}
\ssectionstar{References}\label{sec:References}
\vspace{-0.3cm}

\noindent
(noted in the text and/or recent, relevant ISRs)

\noindent
\href{https://ui.adsabs.harvard.edu/abs/2022AJ....163...78A/abstract}{Ayres, T. 2022}, AJ, 163, 78, On the Same Wavelength as the Space Telescope Imaging Spectrograph

\noindent
\href{https://ui.adsabs.harvard.edu/abs/2019AJ....158..211B/abstract}{Bohlin, R. C. et al. 2019}, AJ, 158, 211, Hubble Space Telescope Flux Calibration. I. STIS and CALSPEC

\noindent
\href{https://ui.adsabs.harvard.edu/abs/2020AJ....160...21B/abstract}{Bohlin, R. C. et al. 2020}, AJ, 160, 21, New Grids of Pure-hydrogen White Dwarf NLTE Model Atmospheres and the HST/STIS Flux Calibration

\noindent
\href{https://www.stsci.edu/files/live/sites/www/files/home/hst/instrumentation/stis/documentation/instrument-science-reports/_documents/2022_07.pdf}{Bohlin, R. C., \& Lockwood, S. 2022}, STIS ISR 2022-07, Update of the STIS CTE Correction Formula for Stellar Spectra

\noindent
\href{https://www.stsci.edu/files/live/sites/www/files/home/hst/instrumentation/stis/documentation/instrument-science-reports/_documents/2018_05.pdf}{Branton, D. 2018}, STIS ISR 2018-05, Performance of the STIS CCD Dark Rate Temperature Correction

\noindent
Brown, T. M. \& Davies, J. E. 2002, STIS TIR 2002-03, Revised Procedures for Creating MAMA P-Flats

\noindent
\href{https://www.stsci.edu/files/live/sites/www/files/home/hst/instrumentation/stis/documentation/instrument-science-reports/_documents/2022_04.pdf}{Carlberg, J. et al. 2022}, STIS ISR 2022-04, Recalibration of the STIS E140M Sensitivity Curve

\noindent
\href{https://www.stsci.edu/files/live/sites/www/files/home/hst/instrumentation/stis/documentation/instrument-science-reports/_documents/2017_06.pdf}{Carlberg, J., \& Monroe, T. 2017}, STIS ISR 2017-06, Updated Time Dependent Sensitivity Corrections for STIS Spectral Modes

\noindent
\href{https://www.stsci.edu/files/live/sites/www/files/home/hst/instrumentation/stis/documentation/instrument-science-reports/_documents/2015_07.pdf}{Cox, C. 2015}, STIS ISR 2015-07, Dark count rates in the STIS FUV MAMA

\noindent
\href{https://www.stsci.edu/files/live/sites/www/files/home/hst/instrumentation/stis/documentation/instrument-science-reports/_documents/200703.pdf}{Dressell, L. et al. 2007}, STIS ISR 2007-03, Time Dependent Trace Angles for the STIS First Order Modes

\noindent
\href{https://www.stsci.edu/files/live/sites/www/files/home/hst/instrumentation/stis/documentation/instrument-science-reports/_documents/199802.pdf}{Ferguson, H. et al. 1998}, STIS ISR 1998-02R, Cycle-7 MAMA Pulse Height Distribution Stability: Fold Analysis Measurement

\noindent
\href{https://www.stsci.edu/files/live/sites/www/files/home/hst/instrumentation/stis/documentation/instrument-science-reports/_documents/200902.pdf}{Goudfrooij, P. et al. 2009}, STIS ISR 2009-02, STIS CCD Performance after SM4

\noindent
\href{https://www.stsci.edu/files/live/sites/www/files/home/hst/instrumentation/stis/documentation/instrument-science-reports/_documents/2024_04.pdf}{Hernandez, S. et al. 2024}, STIS ISR 2024-04, Updating the Sensitivity Curves of the STIS Echelles (Post-SM4)

\noindent
\href{https://www.stsci.edu/files/live/sites/www/files/home/hst/instrumentation/stis/documentation/instrument-science-reports/_documents/2014_02.pdf}{Holland, S. T. et al. 2014}, STIS ISR 2014-02, The Time-Dependent Sensitivity of the MAMA and CCD Long-Slit Gratings 

\noindent
Mason, E. 2013, STIS TIR 2013-02, Revised version of the python code for the creation of STIS CCD P-flats

\noindent
\href{https://www.stsci.edu/files/live/sites/www/files/home/hst/instrumentation/stis/documentation/instrument-science-reports/_documents/2017_04.pdf}{Peeples, M. 2017}, STIS ISR 2017-04, On the Fading of the STIS Ultraviolet Calibration Lamps

\noindent
\href{https://www.stsci.edu/files/live/sites/www/files/home/hst/instrumentation/stis/documentation/instrument-science-reports/_documents/2022_02.pdf}{Prichard, L. 2022a}, STIS ISR 2022-02, STIS CCD \& MAMA Full-field Sensitivity \& Its Time Dependence

\noindent
\href{https://www.stsci.edu/files/live/sites/www/files/home/hst/instrumentation/stis/documentation/instrument-science-reports/_documents/2022_03.pdf}{Prichard, L. 2022b}, STIS ISR 2022-03, Comparison of STIS CCD CTI Corrections on Photometry

\noindent
\href{https://www.stsci.edu/files/live/sites/www/files/home/hst/instrumentation/stis/documentation/instrument-science-reports/_documents/2013_04.pdf}{Proffitt, C. et al. 2013}, STIS ISR 2013-04, Summary of the Results of STIS SMOV4 Calibration Activities

\noindent
\href{https://www.stsci.edu/files/live/sites/www/files/home/hst/instrumentation/stis/documentation/instrument-science-reports/_documents/2017_01.pdf}{Proffitt, C. 2017}, STIS ISR 2017-01, Status of the STIS Instrument Focus

\noindent
\href{https://www.stsci.edu/files/live/sites/www/files/home/hst/instrumentation/stis/documentation/instrument-science-reports/_documents/2017_05.pdf}{Riley, A. et al. 2017}, STIS ISR 2017-05, STIS CCD Performance through Cycle 24

\noindent
\href{https://www.stsci.edu/files/live/sites/www/files/home/hst/instrumentation/stis/documentation/instrument-science-reports/_documents/2015_09.pdf}{Sana, H. et al. 2015}, STIS ISR 2015-09, Summary of the STIS Cycle 21 Calibration Program

\noindent
\href{https://www.stsci.edu/files/live/sites/www/files/home/hst/instrumentation/stis/documentation/instrument-science-reports/_documents/2024_02.pdf}{Siebert, M. et al. 2024}, STIS ISR 2024-02, Recalibration of Pre-SM4 STIS Echelle Throughputs

\noindent
\href{https://www.stsci.edu/files/live/sites/www/files/home/hst/instrumentation/stis/documentation/instrument-science-reports/_documents/2022_01.pdf}{Ward-Duong, K. et al. 2022}, STIS ISR 2022-01, Long-term Rotational Evolution of the STIS CCD Flatfields 

\noindent
\href{https://www.stsci.edu/files/live/sites/www/files/home/hst/instrumentation/stis/documentation/instrument-science-reports/_documents/2018_04.pdf}{Welty, D. E. 2018}, STIS ISR 2018-04, Monitoring the STIS Wavelength Calibration: MAMA and CCD First-Order Modes

%%% Appendix A %%%%
\clearpage
%\vspace{-0.3cm}
\ssectionstar{Appendix A -- Calibration lamps}\label{sec:AppA}
%\vspace{-0.3cm}
STIS has four kinds of internal calibration lamps: tungsten (used for various CCD programs), deuterium (used for NUV-MAMA flat fields), krypton (used for FUV-MAMA flat fields), and Pt/Cr-Ne hollow cathode (LINE, HITM1, HITM2; used for dispersion monitors, slit wheel repeatability).
All of the lamps (except possibly the tungsten lamp) have faded over time (Fig.~\ref{fig:fmflat}; Fig.~\ref{fig:ratios3}; see \href{https://www.stsci.edu/files/live/sites/www/files/home/hst/instrumentation/stis/documentation/instrument-science-reports/_documents/2017_04.pdf}{STIS ISR 2017-04}, \href{https://www.stsci.edu/files/live/sites/www/files/home/hst/instrumentation/stis/documentation/instrument-science-reports/_documents/2018_04.pdf}{STIS ISR 2018-04}, and the plots available on the \href{https://www.stsci.edu/hst/instrumentation/stis/performance/monitoring}{monitors web page}); the spectral shape of the deuterium lamp also appears to have changed.
These changes in lamp behavior have necessitated changes in the exposure time, lamp current, wavelength setting, and/or specific lamp used for some of the calibration observations.
The STIS team is also recommending significant increases in the default exposure times for the automatic wavecals (normally obtained with most STIS spectroscopic observations), for observations at the shortest FUV wavelengths through the narrowest apertures, in order to ensure accurate wavelength zero points; see \href{https://www.stsci.edu/contents/news/stis-stans/august-2023-stan#article2}{2023 August STAN}.

\begin{figure}[!h]
  \centering
  \includegraphics[width=170mm]{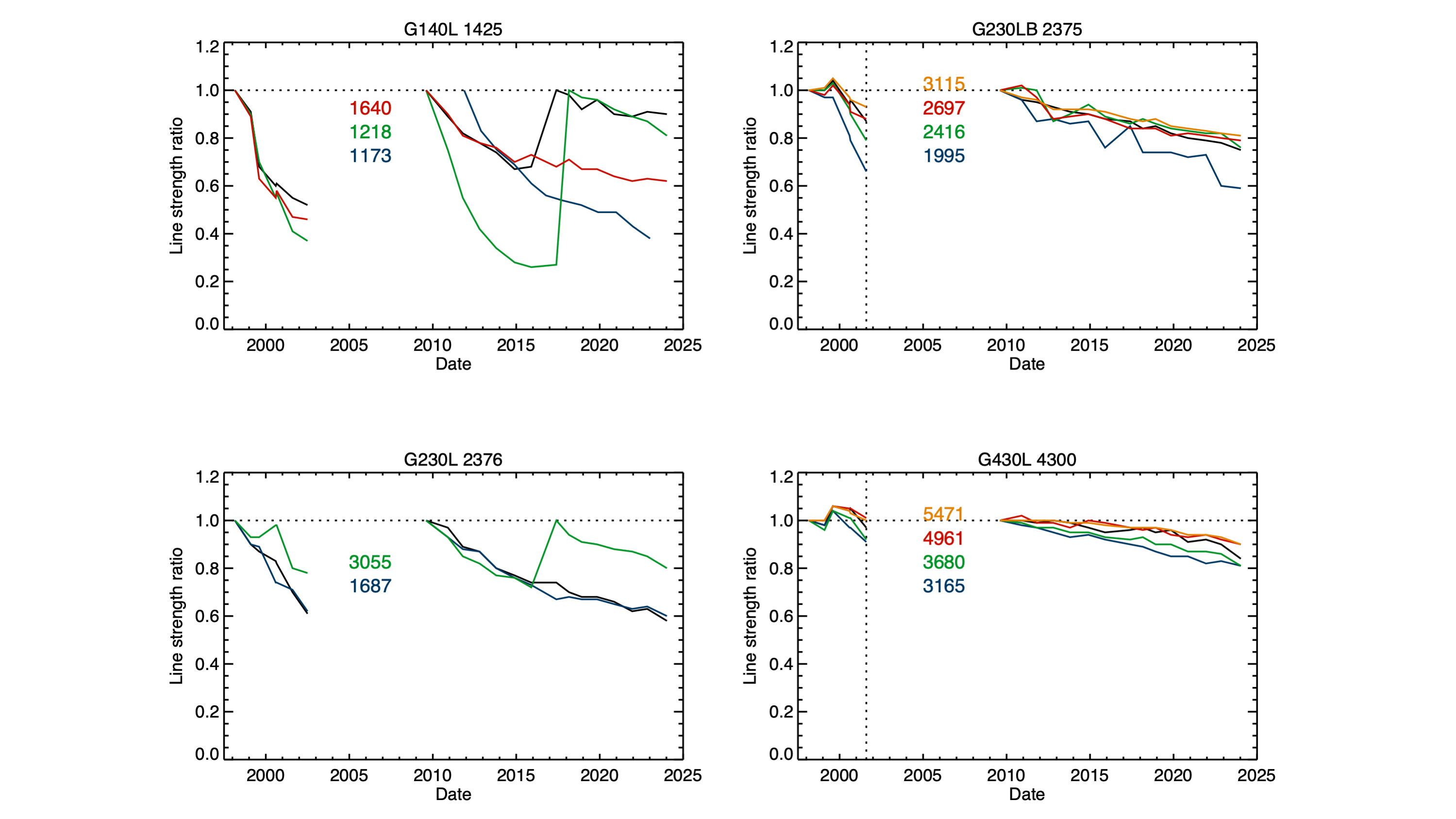}
    \caption{Relative strengths of the emission lines in the Pt/Cr-Ne lamps used for wavelength calibration, for selected first-order G140L/M (FUV-MAMA), G230L/M (NUV-MAMA), G230LB/MB (CCD), and G430L/M (CCD) settings.  Ratios before 2003 are relative to the initial Cycle 7 line strengths.  Ratios after 2009 (post-SM4) are relative to Cycle 17.  Other jumps in the ratios indicate where the lamp and/or the lamp current was changed to obtain more measurable lines.  All three of the Pt/Cr-Ne lamps are fading, most severely at the shortest wavelengths.}
    \label{fig:ratios3}
\end{figure}

%%% Appendix B %%%%
\clearpage
%\vspace{-0.3cm}
\ssectionstar{Appendix B -- Flux recalibration}\label{sec:AppB}
%\vspace{-0.3cm}
Due to recent improvements to the stellar atmospheric models for the primary flux standard stars (\href{https://www.stsci.edu/hst/instrumentation/reference-data-for-calibration-and-tools/astronomical-catalogs/calspec}{CALSPECv11}; see \href{https://ui.adsabs.harvard.edu/abs/2020AJ....160...21B/abstract}{Bohlin et al. 2020}), as well as the re-examination of the Vega spectral flux, the STIS instrument team has undertaken an extensive (and still ongoing) effort to update the flux calibration for all affected STIS modes.
For the low-resolution spectroscopic (L) modes, the improvements to the models of the primary standard stars increase their fluxes by $\sim$2\% between 1500 and 4000 \AA\ and by $\sim$$<$1\% for wavelengths $>$ 4000 \AA, compared to the CALSPECv4 and v5 models that had previously been used to derive the STIS sensitivities.
For the echelle modes, the new CALSPECv11 models increase the FUV fluxes by $\sim$1\% at 1150 \AA\ and by $\sim$$<$3\% at 1700 \AA, and increase the NUV fluxes by $\sim$3\% at 1700 \AA\ and by $\sim$2\% at 3000 \AA.  
In addition to the revised spectroscopic sensitivities, the STIS team has improved the spectral traces for a number of the secondary echelle settings, added calibrations for several ``edge'' echelle orders of particular interest, improved the blaze functions and blaze shift coefficients for some echelle modes, and improved the photometric calibration of the NUV imaging modes.
Several sets of updated reference files (PHOTTAB, IMPHTTAB, RIPTAB, SPTRCTAB) have been released (and data in the MAST archive reprocessed), starting with the modes that are highest priority / most used by the community.
The \href{https://www.stsci.edu/hst/instrumentation/stis/flux-recalibration}{STIS flux recalibration} web page lists the various releases, with links to the STAN articles describing the particulars of each one; see also \href{https://www.stsci.edu/files/live/sites/www/files/home/hst/instrumentation/stis/documentation/instrument-science-reports/_documents/2022_04.pdf}{STIS ISR 2022-04}, \href{https://www.stsci.edu/files/live/sites/www/files/home/hst/instrumentation/stis/documentation/instrument-science-reports/_documents/2024_02.pdf}{STIS ISR 2024-02}, and \href{https://www.stsci.edu/files/live/sites/www/files/home/hst/instrumentation/stis/documentation/instrument-science-reports/_documents/2024_04.pdf}{STIS ISR 2024-04}.
 
\begin{figure}[!h]
  \centering
  \includegraphics[width=170mm]{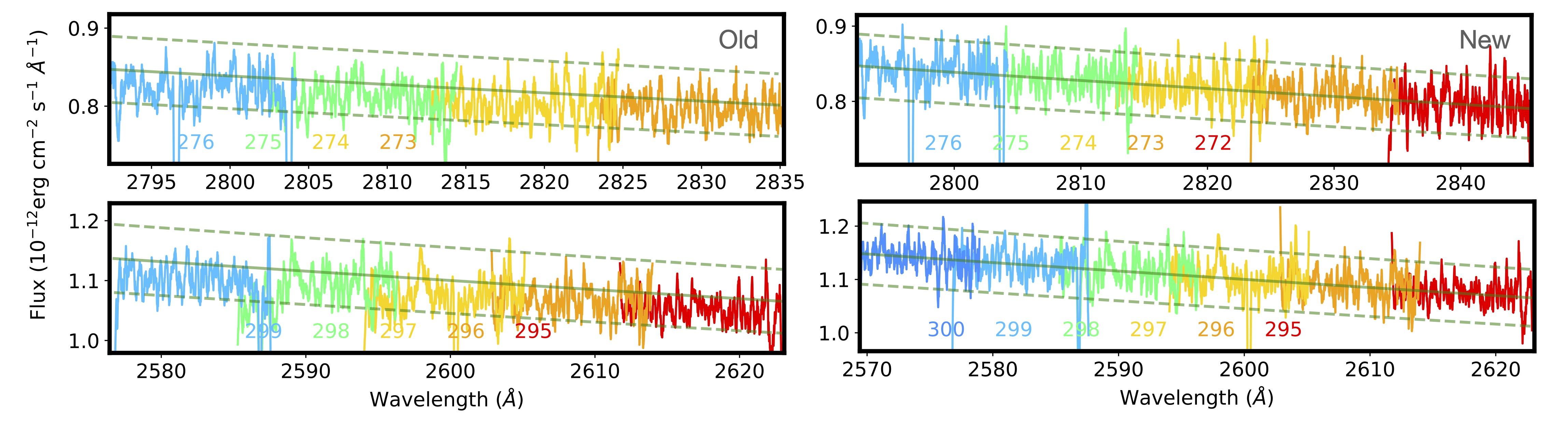}
    \caption{Calibrated E230H/2713 observations for the standard G191-B2B. The corresponding order numbers are shown towards the bottom of the spectra. The green solid line displays the CALSPECv11 model for this standard star, and the dashed lines show the $\pm$5\% intervals. The left-hand panels show the calibrated products using the old PHOTTAB and RIPTAB files. The right-hand panels show the calibration using the recently delivered reference files, better aligned to the CALSPECv11 fluxes. We also include the additional newly calibrated orders 272 and 300.}
    \label{fig:fluxrecal}
\end{figure}

\end{document}